\documentstyle[twoside,fleqn,espcrc2,epsf]{article}
\let\useblackboard=\iftrue
\let\includefigures=\iftrue
\newcommand{\eqa}{\begin{eqnarray}}
\newcommand{\ena}{\end{eqnarray}}
\newcommand{\eqar}{\begin{array}}
\newcommand{\enar}{\end{array}}
\newcommand{\eqn}{\begin{equation}}
\newcommand{\enq}{\end{equation}}

\newcommand{\no}{\nonumber}

\useblackboard
\font\blackboard=msbm10 scaled \magstep1
\font\blackboards=msbm7
\font\blackboardss=msbm5
\newfam\black
\textfont\black=\blackboard
\scriptfont\black=\blackboards
\scriptscriptfont\black=\blackboardss
\def\Bbb#1{{\fam\black\relax#1}}
\else
\def\Bbb{\bf}
\fi
\def\yboxit#1#2{\vbox{\hrule height #1 \hbox{\vrule width #1
\vbox{#2}\vrule width #1 }\hrule height #1 }}
\def\fillbox#1{\hbox to #1{\vbox to #1{\vfil}\hfil}}
\def\ybox{{\lower 1.3pt \yboxit{0.4pt}{\fillbox{8pt}}\hskip-0.2pt}}
\def\comments#1{}

\def\BR{\Bbb{R}}
\def\BZ{\Bbb{Z}}
\def\p{\partial}
\def\half{{1\over 2}}
\def\Tr{{\rm tr\ }}
\def\tr{{\rm tr\ }}

\def\Im{{\rm Im\hskip0.1em}}
\def\bra#1{{\langle}#1|}
\def\ket#1{|#1\rangle}
\def\vev#1{\langle{#1}\rangle}

\def\CH{{\cal H}}

\def\CR{{\cal R}}

\def\ad#1#2{{\delta\over\delta\sigma^{#1}(#2)}}

\def\cint{\oint}
\title{\vbox{\hbox{\rightline{\rm\small RU-94-72}}
\hbox{Large $N$ Gauge Theory --
Expansions and Transitions}}}

\author{Michael R. Douglas\thanks{
Supported in part by DOE grant DE-FG05-90ER40559, NSF PHY-9157016 and the A. J.
Sloan Foundation.}\\
Dept. of Physics and Astronomy\\
Rutgers University\\
Piscataway, NJ 08855
        }
\begin{document}

\begin{abstract}
We use solvable two-dimensional gauge theories to illustrate the issues in
relating large $N$ gauge theory to string theory.
We also give an introduction to recent mathematical work which allows
constructing master fields for higher dimensional large $N$ theories.
We illustrate this with a new derivation of the Hopf equation governing the
evolution of the spectral density in matrix quantum mechanics.

Based on lectures given at the 1994 Trieste Spring School on String Theory,
Gauge Theory and Quantum Gravity.
\end{abstract}

\maketitle

\section{Introduction}

The idea that QCD can be reformulated as some sort of string theory refuses to
die.  If we allow ourselves a sufficiently liberal definition of `string
theory' -- I will take this to be `a theory of embeddings of two-dimensional
manifolds into space-time with an action local both on the world-sheet and in
space-time,' allowing other degrees of freedom as well and noting that the
possible theories have not yet been classified -- we should agree that the idea
has not been disproven.  Nevertheless it has been difficult to make progress
with it.

In these lectures I will describe some of the work which has been done in the
last two years on two dimensional gauge theory.  Although this is drastically
simpler than the four dimensional theory and can only illustrate a few
qualitative features common to both cases, some important ideas have come out
of this work, which suggest new lines to pursue in four dimensions.

A good starting point for discussing `QCD string' is the strong coupling
expansion, which I will review very briefly.  (See \cite{id};
the large $N$ case was discussed in detail by V. Kazakov and I. Kostov at the
1993 Trieste Spring School \cite{kazakov,kostov}.)
We write the path integral
\eqn
Z = \int DA~ e^{- N\beta S_{YM}[A]},
\enq
expand in a power series in $\beta=1/g^2$, integrate termwise, and sum.
To do this, it is necessary to be able to make sense of the functional measure
$DA$ without benefit of the Boltzmann weight, and so far the only way known to
do this in $D>2$ is to latticize the theory.
We thus use the Wilson action $S[U] = \sum_P \Tr U(P)+U^+(P)$ and link
integrals such as
\def\M#1#2#3{#1{}^{i_#2}_{j_#3}}
$\int dU(L)~ \M{U(L)}11 \M{U(L)^+}22 = {1\over N}\M{\delta}12\M{\delta}21$
to evaluate the $O(\beta^n)$ term as a sum over diagrams with $n$ plaquettes
glued together in a continuous and `surface-like' way.
This expansion has a finite radius of convergence (because the number of
diagrams only grows exponentially in $n$) and clearly produces area law
behavior for Wilson loops at small $\beta$.

At large $N$, one can interpret the rules for building diagrams in a way which
makes a diagram with weight $N^{2-2g}$ into a continuous two-dimensional
manifold of genus $g$.
Basically, this is because each link integral, contributing an edge to the
diagram, produces $1/N$; the usual large $N$ coupling dependence gives each
plaquette an $N$, while each vertex comes with a sum over an independent index,
giving $N$.  (The roles of vertices and faces are switched from the weak
coupling expansion.)
The main subtlety is dealing with link integrals such as
\eqa\label{linkint}
&\int dU~ \M U11 \M U22 \M{U^+}33 \M{U^+}44 =
{1\over N^2}\M{\delta}13\M{\delta}31\M{\delta}24\M{\delta}42 \\
&\qquad- {1\over N^3}\M{\delta}13\M{\delta}32\M{\delta}24\M{\delta}41  \qquad\\
&\qquad+ 3\leftrightarrow 4. \qquad\no
\ena
The subleading terms in this expression can contribute to a diagram with
overall weight $N^2$, and thus we must interpret an appearance of this term as
a geometric object which can be glued to the rest of the surface to form a
sphere.
This will work if we call it a `disk' with four edges and zero area, as in
figure 1 -- the geometrical $N$ counting is $N^{1-4}$ which agrees with the
result from the integral.  The minus sign is to be interpreted as a weight
associated with this feature.
Higher `saddles' with more edges also appear in the full expansion.

\includefigures\begin{figure}[bt]
\epsfxsize=3.5cm \epsfysize=5cm
\epsfbox{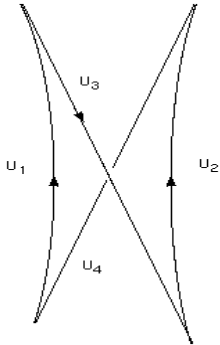}
\caption{A world-sheet `saddle' associated with terms like (3).
Every edge is embedded in the same lattice link.}
\end{figure}\fi

This already looks pretty close to the definition of `string theory' I started
with and thus the project of finding a world-sheet action associated with some
continuum limit of this to get a `gauge string' seems very well motivated.
Already on this qualitative level, many differences with fundamental string
theory are apparent.  Perhaps the most striking is that the world-sheets are
continuously embedded in the target space: this is incompatible with a
world-sheet action $\int d^2\sigma (\p X)^2$ and not easy to describe with a
unitary world-sheet theory, but on the other hand it may explain why in QCD
string theory (unlike the fundamental string) we can define correlation
functions of operators local in target space.~\cite{dqcd}

A serious objection can be raised to starting from a strong coupling expansion:
to take the continuum limit of the theory, we must take
$g\rightarrow 0$ and the lattice spacing to zero as prescribed by the RG, and
surely it is unrealistic to hope that the expansion will have infinite radius
of convergence.  One response to this objection is to argue that our string
theory will agree with gauge theory only at long distances, and that this is an
acceptable limitation.  I disagree with this response.  On theoretical
grounds, if this is what we want, we can simply use the effective long-distance
theory defined by Polchinski and Strominger \cite{ps}, because no structure
specific to the underlying gauge theory survives in the long distance limit.
On practical grounds, what is lacking is a precise calculational technique
which works for intermediate scales, interpolating between the known short and
long distance behaviors, and if we do not get the short distance behavior
right, it is not likely we will get the interpolation right.

Thus we either must abandon the strong coupling expansion or be optimists and
look for a version for which summation and analytic continuation produces the
correct result as $g\rightarrow 0$.
If we can represent it as a string theory, an appropriate treatment of that
theory may well do the analytic continuation for us.
A solid result which speaks against this possibility is the roughening
transition -- I refer to \cite{kogut} for a discussion of this but note only
that the real cause of this was the lattice discretization, and that a
continuum string theory will not have it.

The most direct and explicit version of this idea would be to develop a
continuum strong
coupling expansion.  In $D>2$ we need a regulator, but in principle any
continuum regulator which allows us to make sense of $\int DA$ without the
Boltzmann weight might be used.  Although it would depend on the choice of
regulator, it would surely be a more natural expansion than the lattice
expansion, and conceivably free of spurious phase transitions.
This is one way to make the concept of `QCD string' precise.
We will make contact with my original definition if we can associate terms with
surfaces, if the weight for each surface is determined by local rules, and if
we can reproduce the weight with a local world-sheet action.

Such a string theory has been developed by D. Gross and W. Taylor for
two-dimensional Yang-Mills theory.~\cite{gt}
They gave local rules for the weights and it can also be defined by a continuum
world-sheet action,
as described by G. Moore in his lectures.~\cite{cmr,horava}

There is no obvious reason that a similar expansion for a regulated higher
dimensional theory could not exist.  The regulator would have to act solely on
the measure and preserve gauge invariance -- thus the likely candidates are
stochastic regularizations.~\cite{stochastic,stoq}
I feel this possibility is the most important lesson to be drawn from the work.

Will it be QCD?  In other words, will the $g\rightarrow 0$ limit reproduce weak
coupling physics?  It is difficult to get at this question using the expansion
itself -- a priori, even if it converges, we will need more and more terms to
get a result as $g\rightarrow 0$.  We expect the
world-sheet theory to be strongly coupled at short distances.~\cite{dqcd}
Conceivably, we
might end up with an exact string reformulation of QCD, in which (say) the
one-gluon exchange graph is uncalculable.

A better approach is to use other methods to study the analyticity of the
partition function.  Combining this information with the known finite radius of
convergence of the strong coupling expansion will show its validity.
Clearly we should start by answering this question in the two-dimensional case,
which we can solve exactly.

\section{Two dimensional Yang-Mills theory}

There are many ways to solve the two dimensional theory, and it is worth doing
it in more than one way, first because the techniques come up elsewhere, and
second because it is conceivable that one of them will provide insight for
higher dimensions.

YM$_2$ is very easy to work with, first because the action $\int d^2x~\Tr F^2$
only depends on the volume form, and second, as pointed out long ago by A.
Migdal, there is a lattice definition which is equivalent to the continuum
theory.  The Boltzmann weight (`heat kernel action') $Z(U,A)$ for a plaquette
of area $A$ and boundary holonomy $U$ is just the continuum YM$_2$ path
integral on the plaquette.
Let the plaquette be the disk with coordinates $(r,\theta)$ and $\pi r^2\le A$;
the $A$ dependence can be determined as Hamiltonian evolution with $r$ as time.
Fixing $A_r=0$ gauge and imposing Gauss' law, the wave function depends only on
the holonomy $U=\exp i\int d\theta_0^{2\pi} A_\theta$.
The Hamiltonian is
\eqn\label{groupham}
H = {g^2\over 2N}\tr \left(U{\partial\over\partial U}\right)^2 \equiv {g^2\over
2N} \sum_a E^a E^a.
\enq
This is the only appearance of $g^2$ and from now on we normally choose units
of length in which $g^2=1$.
$E^a = \tr t^a U d/dU$ generates left rotations of $U$ and represents the
Lie algebra $u(N)$.  Thus acting on a wave function which could be any matrix
element of an irreducible representation $R$, $\psi(U)=D^{(R)}_{i\bar j}(U)$,
$H = C_2(R)$, the second Casimir (normalized so that $C_2(\ybox)=N$).
Gauge invariant wave functions $\psi(U)= \psi(gUg^{-1})$
are class functions, which can be expanded in characters
$\chi_R(U) = \tr D^{(R)}(U)$.
Let $\chi_R(U)$ be the character of the irreducible representation $R$ with
$\chi_R(1)=\dim R$; the measure $\int DU$ is normalized so $\int DU~\chi_R(U)
\chi_S(U^+) = \delta_{RS}$.
These results combine to determine the path integral $Z(U_1,U_2;A)$ on a
cylinder of area $A$ and boundary holonomies $U_1$ and $U_2$: the final
ingredient is the boundary condition $Z(U_1,U_2;0)=\delta(U_1,U_2)=\sum_R
\chi_R(U_1) \chi_R(U_2^+)$ on class functions, so
\eqn\label{hkernel}
Z(U_1,U_2;A) = \sum_{R\in \hat G} \chi_R(U_1) \chi_R(U_2^+) e^{-{A\over 2N}
C_2(R)}.
\enq
($\hat G$ is the set of irreducible representations of $G$).
We can then set $U_2=1$, the holonomy for a zero area plaquette, to get
$Z(U;A)$.  By writing $\chi_R(UV)=\Tr D^{(R)}(U) D^{(R)}(V)$ and using
$\int dU D^{(R)}_{i\bar j}(U) D^{(S)}_{k\bar l}(U^+)=
{1\over\dim R}\delta_{RS}\delta_{i\bar l}\delta_{k\bar j}$
one can check the `self-reproducing' property
\eqa
\int dV Z(UV^{-1};A_1) Z(VW;A_2) \no\\
\qquad = Z(UW;A_1+A_2).
\ena
This implies that the lattice integral is invariant under subdividing the
discretization and thus this is equivalent to the  continuum limit.

\def\genus{{\rm g}}
Perhaps the simplest result is the calculation of the partition function on a
Riemann surface due to A. Migdal and B.~Rusakov.
\cite{rusakov}
We can form a genus $\genus$ surface by identifying the edges of a
$4\genus$-gon as $A_1B_1A_1^{-1}B_1^{-1}A_2B_2A_2^{-1}B_2^{-1}\ldots$ and it is
an entertaining calculation to check that
\eqn\label{partg}
Z_\genus(A) =
  \sum_{R\in \hat G} (\dim R)^{2-2\genus}~
  e^{- {A \over 2 N} C_2(R)}.
\enq

We specialize this to the group $U(N)$ to study the large $N$ limit.
This choice turns out a bit simpler than $SU(N)$ but one might worry that we
are introducing extra degrees of freedom which will confuse matters.
The $U(1)$ factor is a gauge theory with gauge coupling $g^2/N\rightarrow 0$ in
the limit, so perturbatively this completely decouples.
However, $U(N)$ is not a direct product but rather a quotient by $\BZ_N$:
one has the exact sequence
$\BZ_N\rightarrow SU(N)\times U(1)\rightarrow U(N)$
where the first map
is $n\rightarrow(e^{2\pi i n/N}{\bf 1},e^{-2\pi i n/N})$
the second is $(e^{i\theta},U)\rightarrow e^{i\theta}U$.
In (\ref{partg}) this means the sum over representations is projected to those
with $U(1)$ charge equal to the $N$-ality of the $SU(N)$ representation, modulo
$N$.  This does affect exact results but in a qualitatively unimportant way.
(It turns out that the $\genus=0$ result is unaffected; this will follow from
the saddle point calculation below and can also be understood from the string
point of view \cite{taylor}).

To get a feeling for it, let us look at the first terms of the series:
\def\euler{{2-2\genus}}
\eqa\label{zser}
&Z_{\genus} = 1 + 2 N^\euler e^{-{A\over 2}} \hskip3cm \no\\
&\hskip 1cm + 2(\half N(N+1))^\euler e^{-{A\over 2}(2+2/N)}\\
&\hskip 1cm + 2(\half N(N-1))^\euler e^{-{A\over 2}(2-2/N)} \no\\
&\hskip 0.5cm + (N^2-1)^\euler e^{-{A\over 2}2} + \ldots \no
\ena
Restoring $g^2$, this is an expansion in $\exp -g^2 A$, a parameter which is
zero in the strong coupling limit.  Thus this is a strong coupling expansion,
and we can study its validity as $g\rightarrow 0$.
This is the limit $e^{-g^2 A}\rightarrow 1$, and this extrapolation already
sounds more attainable than $1/g^2\rightarrow\infty$.  This type of improvement
is common in character expansions.

As was explained here by G. Moore, the work of Gross and Taylor \cite{gt} shows
us how to associate the $N^{2-2G} e^{-nA/2}$ terms in the $1/N$ expansion of
(\ref{zser}) with $n$-fold branched covers of the original surface by genus $G$
world-sheets.  Let us look at $\genus=0$ and $G=0$: the $O(e^{-A/2})$ terms are
easily associated with single covers, while the $O(e^{-A})$ terms in $Z$ are
\eqn
(2N^4 + (\half A^2 - 2A -1)N^2 + O(N^0))e^{-A}
\enq
The first term is the disconnected piece from the $e^{-A/2}$ diagrams.
The $O(N^2)$ terms must be reproduced by double covers of the sphere, and the
rather non-trivial weight can only be reproduced by defining two types of
branch points, which we can call `ordinary' (coming with a power of $A$) and
`$\Omega$-points' (with weight $1$).  This suffices at higher orders, and one
consequence of this is that the $O(e^{-nA/2})$ term associated with $n$-fold
covers is weighted with a polynomial in $A$ with order the number of branch
points, $2n-2$.

The expansion has a very different character for $\genus=0$, $\genus=1$ and
$\genus>1$.  At fixed order in $1/N$, for $\genus>1$ we have finitely many
terms, and no subtleties arise.
For $\genus=1$ one can express the results in terms of modular forms
\cite{dcft,rudd} and the expansion is valid as $g^2 A\rightarrow 0$.

The situation at $\genus=0$ is not immediately clear.  There will be an
$O(N^2)$ free energy, as is true in higher dimensions.  This is the term which
we will relate to genus zero string world-sheets in a string interpretation.
We thus expect it to be `classical' in the same sense that tree diagrams in
quantum field theory are calculable using classical field theory.
This is a central observation and underlies many of the other formalisms which
have been proposed to solve large $N$ field theory; we will come back to it
again.

Let us now reformulate (\ref{partg}) in a way which will provide exact results.
We just need to make the group representation theory explicit, and rather than
do this abstractly, let us find the eigenfunctions of $\Tr E^2$ directly,
i.e. solve the quantum mechanics defined by (\ref{groupham}), following
\cite{dcft}.
(This is also a standard mathematical approach, as in \cite{gurarie,HT}.)

Since we are most interested in class functions, we change coordinates on the
group manifold to $U_{ij}=g_{ik} z_k g^{-1}_{kj}$.
The invariant volume element in these variables is
\eqn
\sqrt{h} = |\Delta(z)|^2 = \tilde\Delta(z)^2
\enq
where $\Delta(z)=\prod_{i<j}(z_i-z_j)$ and
$\tilde\Delta(z)=\prod_{i<j}\sin{\theta_i-\theta_j\over 2}
=\Delta(z)/\prod_i z_i^{(N-1)/2}$.
The ``radial'' components of the metric are simply $h_{ij} = \delta_{ij}$.
Thus on wave functions independent of $g$
\eqn\label{quham}
H = -\sum_i {1\over\tilde\Delta^2}{d\over d\theta_i}
\tilde\Delta^2{d\over d\theta_i}.
\enq
We can rewrite this as
\eqn
H = -\sum_i \left[ {1\over\tilde\Delta}{d^2\over d\theta_i^2}
\tilde\Delta - {1\over\tilde\Delta}
\left({d^2 \tilde\Delta\over d\theta_i^2}\right)
\right] .
\enq
The second term, after some calculation, is found to equal $-N(N^2-1)/12$.

Thus, we can redefine the wave functions by
\eqn\label{waveredef}
\psi=\tilde\Delta\chi,
\enq
and arrive at a theory of $N$ free fermions on the circle.  The boundary
conditions
are also determined by this redefinition; they become periodic (antiperiodic,
respectively) for $N$ odd (even).
An orthonormal basis for wave functions is Slater determinants
\eqn
\psi_{\vec n} = \det_{i,j} z_i^{n_j}
\enq
with energy $E = \sum_i n_i^2 - N(N^2-1)/12$.
The ground state has fermions distributed symmetrically about $n=0$, and energy
zero, so the Fermi level $n_F = (N-1)/2$.

Going back to the original wave functions, we have rederived the Weyl character
formula:
\eqn
\chi_{\vec n}(\vec z) =
{\det_{1\le i,j\le N} z_i^{n_j} \over \det_{1\le i,j\le N} z_i^{j-1-n_F}}.
\enq
In terms of roots and weights, the indices $n_i$ with $n_1>n_2>\ldots>n_N$
are the components of the highest weight vector shifted by half the sum of the
positive roots (usually denoted $\mu+\rho$) where the basis of the Cartan
subalgebra is just $(H_i)_{jk} = \delta_{ij}\delta_{jk}$.
In the language of Young tableaux, if $h_i$ is the number of boxes in the
$i$'th row, $n_i = (N-1)/2 + 1 - i + h_i$.

The $U(1)$ charge is $Q = \sum_i n_i$.  We can change this by a multiple of $N$
by shifting all the fermions $n_i\rightarrow n_i+a$, but $Q\bmod N$ is
correlated with the conjugacy class of the $SU(N)$ representation.

We will do more with this later, but for now we simply adopt the labelling
scheme $\{n\}$ for representations, the formula $C_2=\sum n^2$,
and finally the dimension formula for a representation, computed by rewriting
the determinants as Vandermondes and using l'Hopital's rule to take the limit:
\eqn
\dim R = \lim_{U\rightarrow 1} \chi_R(U) =
\prod_{i>j} \left({n_i - n_j \over i-j } \right).
\enq

Substituting into (\ref{partg}) we find
\eqa
\lim_{N\rightarrow\infty}
Z_\genus(A) = \exp[N^2 F(A)] = \no\\
 ~~ \sum_{n_i\ne n_j} \prod_{i>j} \left({n_i - n_j \over i-j }
\right)^{2-2\genus}
  \exp{- {A \over 2 N} \sum_{i=1}^{N} n_i^2}.
\ena

As pointed out by Rusakov \cite{rusakov},
for $\genus=0$ this expression is such that the sum over $\{n_i\}$ can be
determined by saddle point: it is
\eqn
\exp[N^2 F(A)] =
  \lim_{N\rightarrow\infty}
  \sum_{n_i\ne n_j} \exp -N^2 S_{eff}(\vec n)
\enq
with
\eqa
S_{eff}(\vec n) =
-{1\over N^2}\sum_{i\ne j} \log\left|{n_i - n_j \over N} \right|
  + \\
   { A \over 2 N} \sum_{i} \left({n_i\over N}\right)^2. \no
\ena
Thus our general expectations that `leading large $N$' $=$ `genus zero string
theory' $=$ `classical theory' are confirmed.

\includefigures\begin{figure*}[t]
\epsfxsize=5in \epsfysize=1.7in
\epsfbox{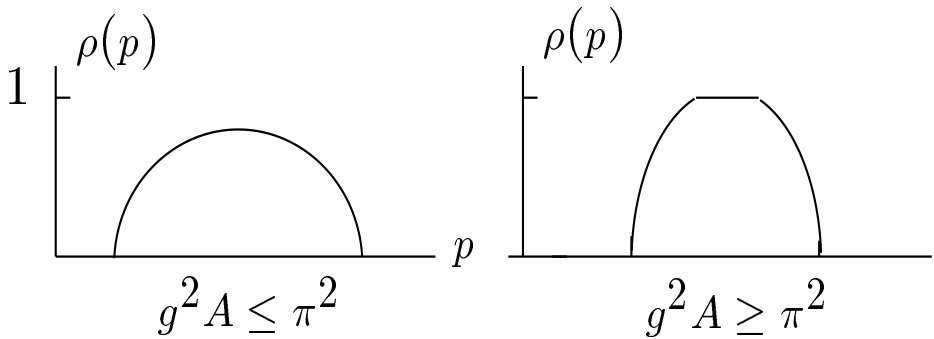}
\caption{Spectral density on $S^2$ in both phases.}
\end{figure*}\fi

$S_{eff}$ contains a repulsive or `entropic' term which favors non-trivial
representations and the sum is exactly the same as a hermitian one-matrix
integral with the eigenvalues replaced by quantized variables $n_i/N$.
Since the quantization is in units of $1/N$ one might expect it not to affect
the leading order saddle point calculation of $F$, and this is almost true.
We thus introduce a scaled variable $p=n/N$ and spectral density
\eqn\label{momdens}
\rho(p) = {1\over N}\sum_i \delta(p-{n_i/N})
\enq
in terms of which
\eqa \label{gzseffsc}
S_{eff}(\rho) =
-\int dp dq ~\rho(p) \rho(q) \log\left|p-q \right| \\
  + {A \over 2 } \int dp~ \rho(p) p^2. \no
\ena

This is the effective action for the simplest of matrix models, the gaussian
integral, which we could just do directly:
\eqn\label{gauss}
\int d^{N^2}M e^{-{N  A\over 2} \Tr M^2} =
\left({2\pi\over N A}\right)^{N^2/2}.
\enq

This is a very simple answer, but it appears to have no relation to the
original sum (\ref{partg})!  The original sum over representations or the
string representation derived from it seems to have been very misleading.
Furthermore, there are independent arguments for the simple answer
(\ref{gauss}), such as the suppression of non-trivial contributions in the heat
kernel on the group manifold evaluated at time $t=A/2N$ \cite{dk} or of
instanton contributions in a direct evaluation of the path integral
\cite{pmgm}.

The resolution of this paradox is that we neglected a crucial consequence of
the discreteness of the variables $n_i$ \cite{dk}: the bound
\eqn\label{momcon}
\rho(p)\le 1.
\enq
The maximum density is attained if the $n_i$ are successive integers.

The saddle point for (\ref{gzseffsc}) satisfies
\eqn\label{oldsad}
{ A \over 2 } p = \int {dq\over p-q} \rho(q)
\enq
and, ignoring (\ref{momcon}) is the semicircle
\eqn
\rho(p) = {A\over 2\pi}\sqrt{4/A-p^2}.
\enq
For $A>A_c=\pi^2$ this violates the bound and we must find a new saddle point,
enforcing the constraint by hand.
Thus, letting $\rho(p)=1$ for $|p|\le b$ and integrating this range
of $p$ in (\ref{oldsad}),
\eqn
{A \over 2 } p - \log {p-b\over p+b}= \int_{-a}^{-b}+\int_b^a~~{dq\over p-q}
\rho(q).
\enq

The general solution of such linear inhomogenous equations is known and
the result is given in \cite{dk}.
It is expressed in terms of elliptic functions, and its series expansion
reproduces (\ref{zser}).
A graph of $\rho(p)$ in the two phases is in figure 2.

For $g^2 A<\pi^2$, the strong coupling expansion fails.
The partition function is non-analytic in a particularly drastic way -- its two
branches are completely unrelated, because `turning on' the constraint is a
non-analytic operation.
This `large $N$ transition' is a consequence of taking the limit
$N\rightarrow\infty$ and does not have a direct analog at finite $N$, as will
be clearer below.

The original result of this type was due to Gross and Witten \cite{gw} and was
generalized by Brezin and Gross \cite{bg} to the matrix integral
\eqn
Z(A) = \int dU e^{N\beta\Tr A U + A^+ U^+}.
\enq
This integral is the generating function for the link integrals
such as (\ref{linkint}) used in deriving the original strong coupling expansion
and the conclusion was drawn that this expansion would be invalid at small $g$.
Now we have learned that the transition is not a lattice artifact.

The double scaled theory around the new transition is the same as that for the
Gross-Witten transition, with the roles of weak and strong coupling
interchanged.~\cite{pmgm}  This may be
intuitively plausible from figure 2.

In terms of the strong coupling expansion, the YM$_2$ transition is signalled
by non-analyticity from summing the polynomial prefactors in (\ref{zser}), and
thus there is a string theory explanation of the transition \cite{taylor}:
the sum over the number of `ordinary' branch points diverges at the transition.

To make a similar argument in higher dimensions, we would need to be able to
control the signs which appear in the expansion.  This is probably not
realistic as it is known that the signs must drastically change the asymptotic
behavior of the string partition function to be subexponential in the
area.~\cite{NielsBohr,dqcd}

All this is a rather serious blow to the QCD string idea as there is no other
convincing argument for an exact string reformulation.
Thus it is essential to understand whether this affects the physics,
whether it persists in $D>2$, and if so what we can do about it.
To say anything about these questions, we need to translate our results to a
formalism which could work in principle in $D>2$.

\section{Collective Fields and Bosonization}

The idea that `the large $N$ limit is classical' can be formalized in many
ways.  At the heart of it is factorization:
\eqn
\vev{\Tr O_1~ \Tr O_2} = \vev{\Tr O_1}\vev{\Tr O_2} + O(N^0).
\enq
This follows directly from the topological expansion -- the factorized term is
a disconnected diagram of $O(N^2)$.
It is as if the functional integral was dominated by a single saddle point --
but one should realize that this is not literally true in $D>2$, and that only
$U(N)$-invariant quantities have well-defined expectation values.
Neither has it been proven beyond perturbation theory in $D>2$,
but its truth in both weak and strong coupling expansions is a rather
convincing argument.

We thus would like to formulate the large $N$ limit as a classical theory whose
configuration space is parameterized by the expectations $\Tr O_i$ and whose
ground state is the solution of some `equation of motion.'  One candidate for
this equation is the factorized Schwinger-Dyson or Migdal-Makeenko equation, as
A. Migdal explained in his lectures:
in the continuum,
\eqa\label{mm}
&\p^\mu(x)\ad{\mu\nu}{x} W(L) =\hfill \\
&\qquad {g^2\over N}\int dy^\nu~
\delta(x-y)~W(L_{xy}) W(L_{yx}) \no
\ena

This equation is quite tractable in two dimensions because of the
area-preserving diffeomorphism invariance.
For present purposes, however, it is easier to work in a canonical formulation,
with which we can make contact with the results of the previous section.

There is a general procedure for finding a `classical' Hamiltonian and phase
space reproducing the large $N$ limit of a general field theory, the collective
field theory of A. Jevicki and B. Sakita.~\cite{jevsak}
The most general and complete treatment of the canonical formalism is due to L.
Yaffe and collaborators, and I highly recommend
\cite{yaffe,yaffetwo}
for the reader's further study.
(We unfortunately did not have time in the lectures to do justice to this.)
The formalism applies in particular to gauge theory in any dimension, and
although there are still not many analytic results from it in $D>2$, there is a
good deal of numerical evidence that it makes sense and properly describes the
large $N$ limit of the regulated theory. (e.g. \cite{yaffetwo,jevicki} and
references there.)

Before plunging into details, let us repeat our primary question:
the strong coupling expansion provides a fairly explicit description
for a candidate ground state of our gauge theory,
so why not just use it?

The simplest derivation of the collective Hamiltonian is to change variables in
the quantum Hamiltonian to invariants.
In $D=1+1$ this is (\ref{quham}), $H =
-{1\over\sqrt{h}}\partial_i \sqrt{h}h^{ij}\partial_j$
and the invariants are
$W_n=\Tr U^n$, or the associated spectral density
$W_n= \int d\theta~ \rho(\theta) e^{in\theta} $.

If we want to write the Hamiltonian using canonically conjugate variables,
we need to make a wave function redefinition like (\ref{waveredef}).
This can be seen by the following heuristic argument:
we want the invariant inner product $(\chi|\psi)=\int d\rho \sqrt{h'}
\chi^*(\rho)\psi(\rho)$ to be simplified by the redefinition
$h'^{1/4}\psi\rightarrow\psi$, so that the self-adjoint momentum operator
$\Pi\equiv h'^{-1/4} (\delta/\delta\rho) h'^{1/4}\rightarrow\delta/\delta\rho$
which has canonical commutation relations with $\rho$.  Carrying this out
produces an `effective potential,' which can be expressed in terms of
invariants.

Let us quote the by now standard result \cite{jevsak,sakita}
\eqn\label{colham}
H_C = {1\over 2N} \int d\theta~ \rho(\partial_\theta\Pi)^2
+ {\pi^2\over 3}\rho^3
\enq
where $\{\Pi(\theta),\rho(\theta')\}=\delta(\theta-\theta')$.
$\rho(\theta)$ is a spectral density and thus there is a constraint
$\int\rho=N$ (which could be implemented with a Lagrange multiplier) as well as
the inequality $\rho\ge 0$.
For now, we will not redefine $\rho\rightarrow \rho/N$,
so we can better explain the relation to the topological expansion as well as
to the original quantum theory.

A simpler set of variables \cite{polchinski} are the chiral combinations
$p_\pm=\partial_\theta\Pi\pm \pi\rho$ satisfying
$\{p_\pm(\theta),p_\pm(\theta')\}=
\pm 2\pi \partial_\theta\delta(\theta-\theta')$:
\eqn\label{colhamtwo}
H = {1\over 12\pi N} \int d\theta~ p_+^3 - p_-^3.
\enq

Classical time evolution under this Hamiltonian reproduces the large $N$ limit
of real time quantum evolution in group quantum mechanics.
The equations of motion derived from (\ref{colham}) are Euler's equations for a
one-dimensional fluid.
(The velocity is $v=\p_\theta\Pi$).
In the variables $p_\pm$ they decouple:
\eqn\label{eulerN}
\partial_t p_\pm + {1\over N}p_\pm \partial_\theta p_\pm = 0.
\enq
This is often referred to as the Hopf equation.
Our YM$_2$ problems are defined in two `Euclidean' dimensions, and we will have
to take this difference into account below: we will see that we should take
$t=i\tau$ and $\Pi\rightarrow -i\Pi$.

In $D=1+1$ one can derive the collective field theory from the simpler free
fermion solution.
(Complete details are given in \cite{dcft}, so we only give a summary in these
notes.)
The invariants $W_n$ are expressed in terms of the fermions as
$\Tr U^n = \sum_i z_i^n$.
We second quantize, introducing the creation and annihilation operators
$\{B^+_m,B_n\}=\delta_{m,n}$ and non-relativistic fermi fields
$\psi(z)=\sum_n B_n z^n$.
We then have $\Tr U^n = \int dz z^{-n-1} \psi^+(z)\psi(z)$.
Thus $\rho(z)=\psi^+(z)\psi(z)$ is the spectral density and we would like to
talk about it as a classical field.

The general answer to this type of problem is bosonization.
However, the classic Coleman-Luther-Mandelstam bosonization can only be applied
to a relativistic fermion.  We can argue as follows in the large $N$ limit:
if we agree not to consider operators $\Tr U^n$ with $|n|\sim N$, we can regard
excitations of the two fermi surfaces as completely independent, and extend the
fermi sea below $n_F=(N-1)/2$ to $n=-\infty$ and similarly extend the sea above
$-n_F$ to $n=\infty$.
The resulting system is kinematically a complex relativistic fermion.
(This is a standard argument in many-body theory.)
The ability to decompose the excitations into right movers $\psi(z)$ and left
movers $\psi(\bar z)$, describing excitations from the two fermi surfaces,
corresponds in the original group theory to decomposing $U(N)$ representations
into a tensor product of a `chiral' representation built from finite tensor
product of fundamentals with an `antichiral' representation built from the
antifundamental.
Rewriting the Hamiltonian
\eqn\label{NRham}
H_{NR} = {1\over N}\int dz~ z^2 \partial\psi^+\partial\psi
\enq
in terms of the relativistic fields produces
\eqa
&H = \cint dz~{z\over 2} :\psi^{+}\partial\psi: + {1\over
N}z^2:\partial\psi^{+}\partial\psi: \\
&+ \cint d\bar z~{\bar z\over 2}:\bar\psi^{+}\bar\partial\bar\psi:
+ {1\over N}\bar z^2 :\bar \partial\bar \psi^{+}\bar \partial\bar \psi: \no\\
&\equiv L_0 + \bar L_0 + {1\over N}H_I
+ {1\over N}\bar H_I \no
\ena
where $L_0$ and $\bar L_0$ are the standard conformal field theory
Hamiltonians.
Applying relativistic bosonization $\psi=:e^{i\phi}:$ produces
\eqn\label{boseham}
H = \cint \half[(\p\phi)^2 +(\bar\p\phi)^2 ]
+ {1\over 3N}[(\p\phi)^3 - (\bar\p\phi)^3].
\enq
This is (\ref{colhamtwo}) with $p_+=N/2+\partial\phi$ and
$p_-=-N/2+\bar\partial\phi$.
(We are taking $\phi$ as a standard free field and thus implicitly defining the
time derivatives in $\partial=\half(\partial_0-\partial_1)$ using free time
evolution generated by $L_0+\bar L_0$.)
We see that it is valid to all orders in $1/N$ (if we attend to quantum normal
ordering).

Let us explain the connection between this formalism and gauge strings.
The strings are the `quantum' fluctuations of the field $\phi$, of $O(1)$.
Since $\Tr U^n = \alpha_n+\bar\alpha_{-n}$ after bosonization,
we see that we reproduce the picture from the strong coupling expansion if we
identify the excitations of the $n$'th left-moving bosonic mode with
$n$-winding strings, and the $n$'th right-moving mode with $-n$-winding
strings.
The leading $O(N^0)$ part of the Hamiltonian (\ref{boseham}) preserves string
number and gives a string energy proportional to the absolute value of its
winding number, while the $O(1/N)$ piece is a three-string interaction.

For present purposes, we can identify world-sheet and target space time, and
think of the world-sheet Hamiltonian as $\int d\sigma \sqrt{X'(\sigma)^2}$.
The complete theory is an interacting `string field theory.'~\cite{MinPoly}
Factors coming from $[\alpha_n,\alpha_{-n}]=n$ make the interaction amplitude
for an $n$ winding string proportional to $n$.
This can be reproduced by a simple vertex which splits or joins strings with
equal amplitude at each target space point.  These are the `ordinary' branch
points in the covariant language.

What makes the discussion intuitively straightforward is the identification of
world-sheet and target space coordinates, making this a tempting assumption in
$D>2$ as well.
However, the canonical approach is not the best way to derive a string because
it loses target space symmetry.
A major advantage of a string reformulation is that what would have been a sum
of diagrams involving any number of vertices, becomes a single diagram in a
covariant approach.

The canonical formulation does allow us to easily compare the string approach
with exact results.
Let us focus on the question: does the genus zero string theory correctly
reproduce all `classical' (leading large $N$) results?
These are $O(N)$ terms in Wilson loop expectations and thus in $\phi$ --
compared to this, the commutators are subleading and thus `quantum.'

If we perturb a coupling by $O(N)$ (or add such a source), the resulting
perturbation of the ground state $\phi$ is also $O(N)$ -- we have changed the
`classical' theory.
Now there is no a priori reason why such results cannot be reproduced by string
theory, even though the assumption that the amplitude of the perturbation is
$O(1)$ is now false.  However there may be `non-perturbative' structure in the
theory, in other words structure which appears only for $O(N)$ shifts of the
fields, and we will miss it in the string theory.

A model which can be solved either by minimizing $H$ or by `string perturbation
theory' is the following, studied by Wadia: \cite{wadia}
\eqn\label{cosham}
H = {g^2\over N} \Tr E^2 + {N\over 2g^2 a^4}\Tr (2 - U -U^+).
\enq

The potential term is a simple approximation to the space-like (`magnetic')
plaquettes of $D>2$ lattice gauge theory -- we could get this model by starting
with the lattice in figure 3, and taking the lattice spacing in the time
direction to zero.
The feature of $D>2$ gauge theory we hope to capture with this is the
following: at very weak coupling, the gauge field
$U(x)=\exp iA(x)$ is roughly Gaussian, and each mode $A(k)$ has a Gaussian wave
function with width $g^2|k|a$.
The potential produces a similar behavior here and we can think of this as the
dynamics of a single mode with $|k|\sim a$.

\includefigures\begin{figure}[bt]
\epsfxsize=6cm \epsfysize=4cm
\epsfbox{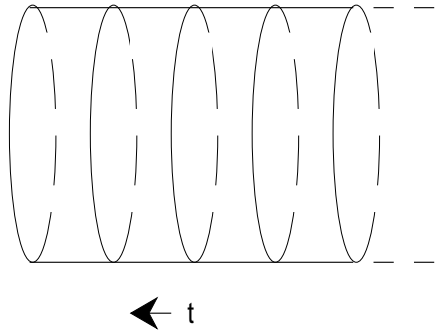}
\caption{A lattice with one plaquette at each time.}
\end{figure}\fi

To find the ground state, we can simply minimize (\ref{colham}) with a
potential term
$\int d\theta~ \rho(\theta)~({1\over g^4 a^4}(1-\cos\theta)-\mu)$
to find the ground state
\eqn\label{cosgs}
\rho = {N\over\pi}\sqrt{\mu-{1\over g^4 a^4}(1-\cos\theta)}
\enq
with $\mu$ determined by $\int\rho=N$.
Again we need to enforce the constraint $\rho(\theta)\ge 0$ on this solution by
hand and there are two cases: for $g^2 a^2>4/\sqrt\pi$ it is never saturated,
while for $g^2 a^2<4/\sqrt\pi$ there is a region around $\theta=\pi$ where it
is.
Thus this model has a large $N$ transition as well.

We could also have expanded $H/g^2$ in $\beta\equiv 1/g^4 a^4$ to get the
Hamiltonian strong coupling expansion, and resummed this expansion.
In the string picture this is a source and one picture we can make is that we
are summing world-sheets with $k$ `holes' at order $1/(ga)^{4k}$, and
integrating over the time in the target for each hole.
Since the source is $O(N)$, terms involving $n$ cubic string interactions and
$n+1$ sources are leading order.

To preserve the $D>2$ analogy, we can also postulate continuous world-sheet
embedded in the `magnetic plaquette.'  It will have different weights from
those in the original expansion: essentially, there are no branch points on the
new world-sheet.  One could enhance the analogy even more by taking the heat
kernel action for the magnetic plaquettes, so the world-sheets would be
generated with the same weights.   This does not change the qualitative
behavior.

Summing all disconnected diagrams will produce a coherent state for the ground
state, of the form
\eqn\label{cohstate}
\ket{W}=\exp\left({\sum_n {W_n\over n} \alpha_{-n}+{W_{-n}\over n}
\bar\alpha_{-n}}\right)~\ket{0}
\enq
with $W_n$ determined from $\rho(\theta)$ in (\ref{cosgs}).

The exact result is analytic in $\beta$ near $0$, so summing the string
diagrams will reproduce it.
If we try to go past the transition by analytic continuation,
we will produce a complex $\rho$, the continuation of (\ref{cosgs}) defined by
simply ignoring the constraints.

Was this a failure of string theory or of the strong coupling expansion?
Really, it is a failure of both.  Built into the free string theory is the
assumption that we can vary the occupation number of each winding number of
string independently.  What we see is that this can fail if the occupation
number or amplitude is $O(N)$.  This is the `non-perturbative structure' we
alluded to above.  We can trace the failure back to the step where we decoupled
the two fermi surfaces, a necessary step in deriving the bosonic theory.

Now the equivalence of (\ref{colham}) and (\ref{NRham}) provides a more general
rewriting the theory in terms of bosonic variables.  We could call it
`non-relativistic bosonization.'  This is particularly interesting in the
present case, since the argument involving decoupling the fermi surfaces broke
down at the turning point (where $\rho$ just reaches zero).
This bosonization is exact in the classical limit, and (with careful treatment
of the turning point) quantum corrections can be derived from (\ref{colham}) as
well.  This is the case relevant for the $c=1$ matrix model.~\cite{cequalsone}

We still cannot use this to fix up our original string theory, because we still
need the constraint $\rho(\theta)\ge 0$ which cannot be expressed in string
language.  What it does suggest is that a broader idea of string theory might
exist, and we will make some comments about this below.

In all the problems we discussed so far, and the one-matrix integrals,
large $N$ transitions all appear exactly where we begin to saturate a
constraint, $\rho(\theta)\ge 0$ or $\rho(p)\le 1$.
Let us see why these are two forms of the same underlying constraint.

Since the eigenvalues of $U$ are classical fermions in the large $N$ limit,
we should be able to specify their positions and momenta simultaneously.
Since they are indistinguishable,
the state of the total system is given by a phase space density
\eqn
\rho(\theta, p) = \sum_i \delta(\theta-\theta_i)\delta(p-p_i).
\enq
Given this, $\rho(\theta)=\int dp~ \rho(\theta, p)$ and $\rho(p)$ of
(\ref{momdens}) is
${1\over N}\int d\theta~ \rho(\theta, p)$.
The fermionic nature of the eigenvalues is exactly captured by
\eqn
0\le\rho(\theta, p)\le {N\over 2\pi}.
\enq
Integrating $d\theta$, we see that the compactness of $\theta$ is responsible
for the upper bound on $\rho(p)$.

The phase space description is more general and in many ways simpler than the
collective field theory. \cite{polchinski}
It can be derived from the quantum (non-relativistic) fermi theory using
\cite{wad}
\eqn
\rho(\theta, p) = \int d\alpha ~ e^{i\alpha p}~
\psi^+(\theta-{\alpha/2N})\psi(\theta+{\alpha/2N}).
\enq
Any one body operator can be written directly as
\eqn\label{onebody}
\sum_i f(\theta_i,p_i) = \int d\theta dp~ f(\theta_i,p_i) \rho(\theta, p).
\enq
(This can be used quantum mechanically and produces Weyl ordered operators.)

For the operators, the large $N$ limit is the standard classical limit.
This is not necessarily true for the states --
assembling $N$ classical fermions will produce states in which $\rho$ at each
point is either $0$ or $\rho_{max}=N/2\pi$, but in general one can construct
states in which
$0<\rho<\rho_{max}$.  An example is the state corresponding to a specific
representation $R$, i.e. with wave function $\det z_i^{n_j}$.
Such states are not coherent states of the form (\ref{cohstate}) and do not
usually come up in practice.

The commutator of operators becomes Poisson bracket, a free particle bracket
$\{f,g\} = 2\pi(\partial_p f \partial_\theta g - \partial_\theta f \partial_p
g)$.
Time evolution is $\dot\rho=\{H,\rho\}$.

Collective field theory is derived by assuming a form
\eqn
\rho(\theta, p) = \cases{{N\over 2\pi}&for $p_-(\theta)\le p\le p_+(\theta)$\cr
0&otherwise.\cr}
\enq
For example, $H={N\over 2\pi}\int dp~ \rho(\theta, p) \half p^2$
reproduces (\ref{colhamtwo}).
Again, the simple form of the state is an assumption, which is easily seen to
be correct for problems such as (\ref{cosham}).
Several generalizations are possible.
One can have a fermi surface of arbitrary shape, and a collective field theory
using several functions $p_i(\theta)$ to describe it.
There is also nothing special about the parameterization $p(\theta)$ and one
could use $\theta(p)$ or $(\theta,p)$ depending on a parameter $s$ to describe
the surface.

Free fermions are special to $D=1+1$ (and this system), but a language which in
principle could generalize some of this to $D>1+1$ is known.~\cite{yaffe}
We can construct the phase space as the orbit of a reference state under the
action of a `coherence group,'  obtained by exponentiating the Lie algebra of
observables.
Thus the entire kinematics of the limiting large $N$
theory is contained in the geometry of this group.

The discussion of \cite{yaffe} started from the basic observables of collective
field theory, $\rho(\theta)$ derived from the invariant operators $\Tr U^n$ and
$\rho\partial\Pi(\theta)$ from $\half\Tr (EU^n + U^nE)$.
These generated a semi-direct product of a Virasoro algebra with the modes of
$\rho(\theta)$ and the corresponding group ${\rm Diff}(S^1)$ semi-direct
product the additive group of functions $\rho(\theta)$.
The algebra generalizes directly to $D>1+1$ and in the next section we will be
able to say something about the coherence group.

More recent work on these lines \cite{wad,sak,cap}
uses all the one-body operators (\ref{onebody}) to generate the group of
symplectic diffeomorphisms on phase space, ``${\rm SDiff}(2)$.''
The idea that the canonical formalism is simpler if we consider all invariant
operators made from $U$ and $E$, not just low powers of $E$, has not been
studied in $D>2$, and might be valuable there as well.

So far, we have seen a relation in $D=1$ and $D=1+1$ between large $N$
transitions and constraints on the invariant observables.
Similar constraints exist in higher dimensions.
Essentially nothing is known about the phase space version, but the analog of
$\rho(\theta)\ge 0$ is clear.
Given any loop $L$ in space-time (or lattice), we can consider its holonomy
$U(L)$ and associated spectral density $\rho_L(\theta)$: this should satisfy
the same constraint.
We will look at this in more detail in the next section.

It is natural to conjecture that large $N$ transitions are always associated
with a change in the application of the constraints.
Now, to have any hope of making a definite statement in higher dimensions, we
need to find some qualitative conditions for the transition, that do not
require exact results.
Thus we might hypothesize that if for coupling $g_1$ the ground state saturates
a constraint, and for $g_2$ it does not, there is necessarily a large $N$
transition in between.

This is an attractive hypothesis because it can be checked at strong and weak
coupling where we have some control over the higher dimensional gauge theory.
For strong bare coupling, $\rho_L(\theta)$ (for a loop on any scale) should
saturate no constraints ($\rho>0~ \forall\theta$), while at weak coupling and
short distances the gauge field fluctuations are highly suppressed and we
expect $\rho_L$ for small $L$ to have support sharply peaked about the origin
(with width $\sim~g^2a/|L|$), no matter what regulator we use.
This has been seen in the numerical studies \cite{jevicki,yaffetwo} and the
finite radius of convergence of both diagram expansions makes it accessible
analytically.
The only doubt we might have regards the weak coupling phase, where to be
precise we must be able to distinguish $\rho(\theta)=0$ from a very small but
nonzero value, but let us imagine we can prove that (say) $\rho(\pi)=0$ for
small loops.

One might think (and at Trieste I advocated the idea) that this would prove the
existence of the large $N$ transition in higher dimensions, and the failure of
the strong coupling expansion.  We might expect $\rho_L$ for small loops to
become complex, like (\ref{cosgs}), which looks very unphysical.

However, there are subtleties in this, as one realizes on considering the
Wilson loop expectations in two dimensions on the plane.
These are determined directly from the Boltzmann weight,
$Z(U;A)=Z(U,1;A)$ as in (\ref{hkernel}):
\eqa\label{scplane}
W_n = \int dU~\Tr U^n~\sum_R \chi_R(U) \dim R e^{-{A\over 2N}C_2(R)} \no\\
= \sum_{m=0}^n (-1)^m \dim R e^{-{A\over 2N}C_2(R)} \no
\ena
\includefigures\begin{figure}[bt]
\epsfxsize=5cm \epsfysize=4cm
\epsfbox{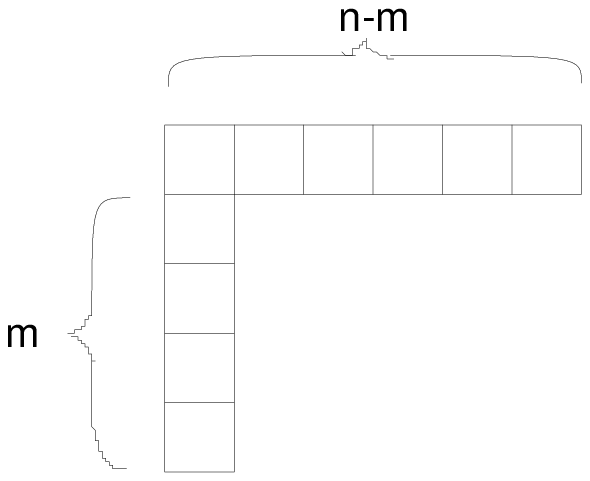}
\end{figure}\fi

The sum is over the representations in the above figure.
(This can be derived from fermionization:
$\Tr U^n\ket{0}=\sum_m b^+_{1/2-n+m} b_{-1/2-m}\ket{0}$.)
Since the sum is finite, this will be analytic in $A$.
(The string diagrams can have up to $n-1$ `ordinary' branch points).
If we compute $\rho(\theta;A)$ from these, we will find that it satisfies the
conditions of the hypothesis:
for $g^2 A < 4$, $\rho(\theta)$ has a gap around $\theta=\pi$, while for
$g^2 A > 4$ it does not.~\cite{durole}
We might call this a `pseudo-transition'.  It is visible in observables, but we
need to look at Wilson loops winding arbitrarily many times to see it:
at fixed $g^2 A < 4$, the $W_n \sim n^{-3/2}$ as $n\rightarrow\infty$, rather
than falling exponentially.

We thus see that non-analyticity in the coupling for individual Wilson loop
expectations and the existence of phase transitions cannot be concluded solely
from the behavior of the spectral density, as pointed out long ago by B.
Durhuus and P. Olesen.~\cite{durole}
Actually this does not directly contradict the hypothesis as stated, because
the states in question are not ground states of the Hamiltonian.
Nevertheless our belief in it may be somewhat shaken.

One can compute $\rho(\theta;A)$ more directly from the collective field
theory.
The problem was actually solved this way in \cite{durole}, without ever
mentioning collective field theory -- they
derived (\ref{euler}) directly from the Migdal-Makeenko equation.

Let us do `radial quantization' on the plane, embedding the Wilson loops at
fixed radius $r$, and and then go to the canonical formalism using $A=\pi r^2$
as time.
Let us also finally redefine $\rho\rightarrow\rho/N$ from (\ref{colham}).
$\rho(\theta;A)$ is thus the result of evolution for Euclidean time $A$ from
the boundary condition $\rho(\theta;0)=\delta(\theta)$.
It is simplest to think of going to Euclidean time in the action
$S = \int dt \Pi\dot\rho -H$, by taking $t=iA$.  To preserve the commutation
relations in the Hamiltonian we should restore the first term by taking
$\Pi\rightarrow -i\Pi$.  It is then convenient to redefine
\eqn
p_\pm=\partial_\theta\Pi\pm i\pi\rho
\enq
The equations of motion are the same (except for the $1/N$):
\eqn\label{euler}
\partial_A p_\pm + p_\pm \partial_\theta p_\pm = 0.
\enq

A nice simplification is now possible: since $\Pi$ and $\rho$ are real, the two
equations are complex conjugate, and
and if $p_+(\theta)$ is analytic, we have reduced the equation to a first-order
ODE.  Since $\rho(\theta)$ and $\Pi(\theta)$ are only prescribed on the real
axis, there is generally no difficulty in finding such an analytic function.

We don't know $\Pi$ at $A=0$ -- it is not determined by the quantum initial
condition $\psi(U)=\delta(U,1)$.
In general, the classical limit of a Euclidean time quantum problem becomes a
boundary value problem, where we specify (say) initial and final positions.
The initial and final momenta are then determined only by solving the equations
of motion.  In general, more than one solution can exist with the same boundary
conditions, and this is a possible source of phase transitions.
All this applies to the large $N$ limit
and in the present case we must specify the limiting behavior of $\rho$ as
$A\rightarrow\infty$: it is the strong coupling vacuum $\rho(\theta)=1/2\pi$.
(Similar considerations apply to the large $N$ limit of the Itzykson-Zuber
integral, which as shown by A. Matytsin, can be expressed in terms of the same
boundary value problem.~\cite{matytsin})

Given only the boundary conditions, the explicit solution is not easy to find.
However there is one more fact about the solution, which greatly simplifies the
problem.
After extending $p_\pm$ to be analytic functions in the complex $\theta$ plane,
they fall off as $p_\pm \sim 1/\theta$ for large $\theta$, much like the
resolvent $\Tr (z-U)^{-1}$.
Why this should be is not at all obvious from what we have said so far,
but this assumption makes the problem easy, because then $\Pi$ at $A=0$ is
determined by a dispersion relation, and we have an initial value problem.
We can check the assumption at the end by verifying that it produces the
correct $A\rightarrow\infty$ limit.

The final ingredient is to take periodicity in $\theta$ into account, which we
can do by periodizing the initial condition.
The analytic function satisfying
$\Im f=\pi\sum_n\delta(\theta-2\pi n)$ is then
\eqn\label{bc}
f(\theta) = {1\over 4\pi} \cot {\theta\over 2}.
\enq

The general solution of the Hopf equation can be derived by first representing
the boundary $p_+(\theta)$ in phase space parametrically as
$(p_+(\xi),\theta(\xi))$, implementing free motion
$\theta(\xi;A)=\theta(\xi;0)+Ap_+(\xi)$, and solving for $\xi$.
Although the pictures are simplest for real $p_+$, the solution is perfectly
valid for complex $p_+$.

Thus, given an initial condition $p_+(\theta;A=0) = f(\theta)$, the solution is
\eqn\label{hopfsol}
p_+(\theta;A) = f(\xi(\theta;A))
\enq
where $\xi$ is the solution of the implicit equation
\eqn\label{hopfsoltwo}
\xi(\theta;A) = \theta - A f(\xi(\theta;A))
\enq
We can then recover $\rho={1\over \pi} \Im p_+ = -{2\over A} \Im\xi$.

This cannot be solved in closed form, but the
comparison of $\rho(\theta)$ with (\ref{scplane}) for $A\rightarrow \infty$,
and the verification of the properties we described are not hard
(and done explicitly in \cite{durole}).  For example, to check that
$\rho(\theta;A)\rightarrow 1/2\pi$ as $A\rightarrow\infty$, we can take the
imaginary part of (\ref{hopfsoltwo}) and check that
$\Im\xi\rightarrow -A/4\pi$.

What is the essential difference between this problem and the sphere partition
function?
In this language, the sphere problem is formally very similar:
it is a two-point boundary problem where both initial and final
boundary conditions are $\rho(\theta)=N\delta(\theta)$.

Again, boundary value problems are not easy to solve, so we ask if another
trick can convert this into an initial value problem.
We can always extend $p_\pm$ into the complex $\theta$-plane as analytic
functions, but now they cannot fall off as $p_\pm\sim 1/\theta$, simply because
$\Pi$ is different in the two problems.
I have not found such a trick and since I just want to illustrate the
qualitative aspects of the problem, let us instead solve the problem for
fermions on the real line.
In other words we replace YM$_2$ by hermitian matrix quantum mechanics.
In this case we already know the solution of the quantum problem with boundary
conditions $\psi(M;t=0)=\delta(M)$: it is $\ket{t}\propto\exp -N\Tr M^2/t$.
Using this we can compute
\eqn
\rho_t(\lambda) = \vev{T-t|{1\over N}\sum_i \delta(\lambda-\lambda_i)|t}
\enq
and it is the saddle point spectral density for
\eqn
\int \prod d\lambda_i~\Delta(\lambda)^2 \exp -N({1\over t}+{1\over
T-t})\sum_i\lambda_i^2.
\enq
This is
\eqn
\rho_t(\lambda) = {2\over\pi r^2}\sqrt{r^2-\lambda^2}
\enq
with $r^2(t)=4t(T-t)/T$.
One can compute $\Pi_t$ similarly and check that the resulting
\eqn
p_+(\lambda;t)= {2\over r^2}\left( (1-{2t\over T})\lambda-
\sqrt{\lambda^2-r^2} \right)
\enq
is a solution of (\ref{euler}).

In fact, this is the weak coupling solution for the YM$_2$ sphere problem as
well (found this way in \cite{cas}), with $T=A$.
The periodicity $\theta\cong\theta+2\pi$ of eigenvalue space is completely
invisible until we reach $r(A/2)=\pi$.
This is the position space version of what we saw in section 2.

Once we reach $A=\pi$, the non-linearity of (\ref{euler}) means this is no
longer a solution.  In the phase space formalism this is due to the constraint
$\rho\le \rho_{max}$.
One could do a calculation similar to that of the Wilson loop on $S^2$
\cite{Boulatov,Daul} to find the full strong coupling solution.

In this language, the difference between the two problems is subtle -- the
plane problem `knows about the compactness of eigenvalue space from the
beginning,' while the sphere problem `does not know until it hits the
constraint.'
It all comes from the boundary conditions:
whereas the sphere problem can be completely taken over to non-compact
eigenvalue space, the final boundary condition for the plane
$\rho(\theta)=1/2\pi$
only makes sense for compact eigenvalue space, and enforcing it affects the
solution at all times.

The compactness enters in a slightly strange way in the initial value
description: although the original problem was free fermions, which a priori
would not know whether they were moving in compact or non-compact space,
when we periodize the boundary condition (\ref{bc}), we change the motion
around $\theta=0$, because (\ref{euler}) is non-linear.

The goal of this section was to study the large $N$ transition entirely in
terms of invariant observables, since this language can generalize to $D>2$,
and allows contact with a string reformulation.
Clearly any conclusions we try to draw for $D>2$ will be speculative, but
let us imagine we have a local continuum regulated gauge theory in $D>2$.
We need a cutoff and thus the dependence of observables on length scale and
coupling constant are not directly related.
We will study it on $\BR^D$.

We can study the dependence on length scale by doing radial
quantization of the theory, which turns scale dependence into `time'
dependence.
Like the YM$_2$ expectations on the plane this should be a classical boundary
value problem, and the large radius boundary condition provides a way for the
system to `know' about the non-trivial topology of configuration space and
avoid a transition as a function of length scale.

On the other hand, the strong coupling expansion has no such mechanism --
it only sees the limit $g\rightarrow\infty$ in which the constraints, the
non-trivial structure of configuration space, are invisible.
Thus summing it will produce results which have no reason to satisfy any
constraints.
The same conclusion would apply to any string reformulation in which
the state with zero strings is the extreme strong coupling state
$W_0(L)=N\delta_{L,I}$.~\footnote{
In other words, a reformulation in which this result for the limit
$g\rightarrow\infty$ is reproduced in the `obvious' way -- by the world-sheet
Boltzmann weights going to zero, as opposed to cancellations between large
numbers of diagrams.}
The true $W(L)$ is produced by
summing string diagrams, and nowhere in this formalism are the constraints
enforced.

YM$_2$ on the plane is a special case because the observables depend only on
$g^2A$ and thus
the first argument eliminates the transition in $g$.~\footnote{
Conceivably, we could derive a string in $D>2$ from a continuum formulation
with no cutoff dependence or directly from the renormalized theory, in which
case this would also apply.}
At present, all we can
infer from this is that the second argument does not prove there will be a
transition.

What about a system in finite volume?  Let us start by describing another way
to understand the transition for $S^2$, developed by D. Gross and A. Matytsin
\cite{pmgm}:
it is driven by non-trivial classical solutions in the path integral.
One can evaluate the path integral for $Z(S^2)$ as a sum over classical
solutions, using a non-abelian generalization of the Duistermaat-Heckman
formula.~\cite{witten}
These are unstable solutions contained in $U(1)^N$ and constructed from the
$U(1)$ instanton.  (There is no $SU(N)$ instanton in $D=2$ as bundles over
$S^2$ with gauge group $G$ are classified by $\pi_1(G)$.)

The $U(1)$ instanton is the configuration Dirac proposed as a monopole in
$D=4$, restricted to constant $r$ and $t$.  As usual for instantons in large
$N$ gauge theory it has $O(N)$ action, from the explicit $N$ in $e^{-NS}$.
However it can survive the large $N$ limit because of `entropy' -- each $U(1)$
factor has an independent instanton number -- and in the strong coupling phase
it does.
This could be thought of as a covariant description of the `windings of
eigenvalues around $S^1$' which would appear if we made the classical fermion
picture discussed here completely explicit.

Gross and Matytsin propose that the $D=4$ transition could also be driven by
topologically non-trivial configurations, the standard instantons.
In \cite{pmgm}, they make a one-loop estimate of their effect as a function of
the volume of the system (which controls the maximum of the running coupling)
and find a result consistent with no phase transition.
(See also \cite{neuberg}).

{}From the reasoning we gave earlier, we would conclude instead that we
generally expect transitions as a function of volume in $D\ge 2$.
At some critical volume, a mode will be strongly enough coupled to explore its
full configuration space (e.g. $0\le\theta\le 2\pi$), new constraints will come
into play, and new solutions of the collective field theory can exist.
Now Gross and Matytsin are careful to state that they are arguing against a
{\it large $N$ instanton induced phase transition} so there is no direct
contradiction between these two statements: the instanton is certainly not the
only field configuration which can explore the full configuration space.

In any case these questions must be considered unsettled, and the main point of
my discussion is to outline considerations which we might be able to make
precise given more control over $D>2$ collective field theory (or other loop
equations).  The main problem for now is to find a continuum regulated theory
we can work with in $D>2$.

So far I have essentially been identifying `gauge string' with `continuum
strong coupling expansion.'
Now this is really an assumption and if it turns out that the strong coupling
expansion is invalidated by a large $N$ or other phase transition, we would
certainly want to change it.

What went into this assumption?
The main element was the assumption that the state (or configuration -- the
discussion is really parallel in the canonical and covariant formulations) with
no strings is the loop functional $W_0(L)=N\delta_{L,I}$.

As we saw, `strings' are a reformulation of perturbation theory,
not necessarily in the coupling, but in the sense that they describe $O(1)$
variations of an $O(N)$ initial configuration.
Thus the choice of initial configuration to expand around is crucial.
There is no way to describe constraints such as $\rho_L(\theta)\ge 0$ if we are
restricted to statements expressable in perturbation theory around the extreme
strong coupling state.
This is a good motivation to change our zeroth order state.

One candidate which comes to mind is the extreme weak coupling state,
$W_1(L)=N$ for all loops.  Thinking about this leads us to ask the question,
can standard weak coupling perturbation theory be reformulated as a theory of
continuum world sheets, if we restrict ourselves to computing gauge-invariant
observables?  This is an interesting question which we will not discuss here.
In any case this perturbation theory is seriously flawed by infrared
divergences coming from the massless gluons.  These do not cancel in all
gauge-invariant quantities and no consensus has emerged as to whether it can be
used even in principle for low energy physics.\footnote{What is better
understood is to introduce an IR cutoff by hand, either by
Higgsing, or by putting the system in a small finite volume in which the
coupling
does not get strong.  The classic study of the large $N$ weak coupling
perturbation theory is \cite{thooft}.  More recently Yang-Mills theory (at
finite $N$) has been shown to be rigorously definable in finite volume.
\cite{mrs}}

Both starting points are handicapped by their serious qualitative differences
with the true vacuum of the theory.
Now in principle, we could imagine taking any loop functional $W_g(L)$ as our
starting point for perturbation theory.  It may not even be necessary for it to
be the solution of some solvable limit of the theory.
Let us write the true solution as $W(L)=W_g(L)+\delta W(L)$, and imagine
deriving an expansion by plugging this into (\ref{mm}).
In general we will get terms of order $(\delta W)^1$, $(\delta W)^2$ and if
$W_g(L)$ does not solve the equation, $(\delta W)^0$.
We will want to interpret the first and second terms as leading to a `string
propagator' and `string vertex' -- and the third term as a `string source,'
which might be modeled as holes in the world sheet, but much better as
additional pieces of world-sheet with an action chosen to produce the correct
dependence on $L$.

We could also change the association of small fluctuations of the loop
functional $W(L)$ with `strings,' as long as we preserve the locality of the
resulting world-sheet theory.

To get new starting points, we need to have some non-perturbative way to
construct loop functionals.  Now the space of loop functionals is sometimes
said to have little structure: one just specifies $W(L)$ for each $L$
satisfying $|W(L)|\le N~~\forall L$.  This is not true and in fact the space is
not easy to describe in $D>1$.  What is easier to describe are master fields.

\section{Master Fields}

Although we argued that taking advantage of factorization required us to
restrict attention to $U(N)$ invariant quantities, in YM$_2$ it was very useful
to talk about the holonomy for a given loop and its spectral density.

The higher-dimensional generalization of this is the `master field,' a single
gauge configuration $A_\mu(x)$ defined to reproduce all Wilson loop expectation
values by evaluation: $W(L)=\Tr P \exp i\int_L A_\mu(x) dx^\mu$.
Since the only assumption is factorization, the idea is also valid for large
$N$ vector and matrix models, and the master field has been worked out for some
vector models.~\cite{jevlev}

Physicists have not had much success in writing
explicit master fields for higher
dimensional gauge theory or matrix models.  It turns out, however, that recent
mathematical work, specifically in the field of operator algebras, allows this
to be done.

The most extensive work has been done by D. Voiculescu and collaborators.
I will only give an introduction here, leaving out many interesting results,
which are described very clearly in \cite{voi}.
More recently, I. Singer has shown how to construct the master field for
YM$_2$.~\cite{singer}

The problem we discuss here is to compute leading order expectation values
under integrals over several hermitian matrices $M_i=M_i^+$.
Let the index range over the set $S$.
Using factorization, the general case is
\eqn\label{mmat}
\vev{W} = {1\over Z} \int~ \prod_{\alpha\in S} d^{N^2}M_\alpha~e^{-N V}
{}~~{1\over N}\tr W
\enq
where $W$ is an arbitrary `word' or product of matrices
\eqa\label{word}
W &= M_{\alpha_1}^{n_1} M_{\alpha_2}^{n_2} M_{\alpha_3}^{n_3} \ldots~\\
&= \prod_{i=1}^m M_{\alpha_i}^{n_i}.\no
\ena

We will restrict ourselves to the case
\eqn\label{decoup}
V = \sum_{\alpha\in S}  \tr V_\alpha(M_\alpha),
\enq
in other words no coupling between the matrices.
Thus expectation values $\vev{\tr M_\alpha^n}$ are determined by the usual
saddle point analysis and spectral densities $\rho_\alpha(\lambda_\alpha)$,
and this is hardly a `$c>1$' matrix model.
Nevertheless we have retained an aspect of the $c>1$ problem:
the total number of words of a given length $l$, $|W_l|$, grows exponentially
with $l$, and it is non-trivial to compute their expectations just given the
$\rho_\alpha(\lambda_\alpha)$.

Still, it is not very hard to do it -- one can consider words with successively
larger $m$ in (\ref{word}), and write a chain of Schwinger-Dyson equations
determining the case $m$ in terms of lower order expectations.
Nevertheless this is rather unwieldy and one can ask for a simpler description
of the result.
This could be provided by a `master field' -- a set of single matrices ``$\hat
M_\alpha=\vev{M_\alpha}$'' which we could regard as the saddle point for the
integral (\ref{mmat}).  Now for each of the individual matrices, there is
certainly a diagonal matrix with this property, but we cannot simply use this
set:
e.g. consider $V=\tr A^2/2+\tr B^2/2$ for which
\eqn
\vev{\tr A^2B^2} = N
\enq
while at leading $O(N)$
\eqn
\vev{\tr ABAB} = 0.
\enq

Now an integral like (\ref{mmat}) is not literally dominated by a saddle point
for the $M_\alpha$.  If we write $M_\alpha = U_\alpha D_\alpha U_\alpha^+$,
clearly fluctuations of the relative angular degrees of freedom $U_\alpha
U_\beta^+$ are completely unsuppressed.
On the other hand, the graphical argument for factorization can be made
rigorous, and it is not clear why the heuristic arguments for the master field
should fail.

The reconciliation of these two statements is that we can contruct a master
field which behaves as if it were the saddle point.
Let us do it for the case $V_\alpha(M_\alpha) = \half \Tr M_\alpha^2$.

We define the `free Fock space on $S$'%
\footnote{In this section we will always use the word `free' in its
mathematical sense, to mean a group or algebra with no non-trivial relations,
rather than its physical sense.}
to be a Hilbert space with orthonormal basis vectors labelled by an ordered
list of zero or more elements $\alpha_j\in S$.  The elementary operators acting
on this are $l_\alpha$ and $l_\alpha^*$ defined as follows:
\eqa
l^*_\alpha \ket{\alpha_n,\alpha_{n-1},\ldots,\alpha_1} &=
\ket{\alpha,\alpha_n,\alpha_{n-1},\ldots,\alpha_1}\hfill\cr
l_\alpha \ket{\alpha_n,\alpha_{n-1},\ldots,\alpha_1} &=
\delta_{\alpha,\alpha_n} \ket{\alpha_{n-1},\ldots,\alpha_1}\hfill\cr
l_\alpha \ket{} = 0.
\ena
They are similar to bosonic creation and annihilation operators but with two
differences.  First, they are {\it free} -- the product of two operators
associated with $\alpha$ and $\beta\ne \alpha$ satisfies no relations.  Second,
the usual symmetry factor for bosonic harmonic oscillators is absent.
Thus, we have not $[l,l^*]=1$ but instead
\eqn
\sum_\alpha [l_\alpha,l^*_\alpha] = \ket{}~\bra{}.
\enq

On this Hilbert space, we define an operator $\hat M_\alpha$ associated with
each $M_\alpha$, and a `trace' $\hat\tr$.
For the gaussian integral $V=M^2/2$,
\eqn
\hat M_\alpha = l_\alpha + l_\alpha^*.
\enq
The `trace' is not the standard one (which does not make sense here)
but is defined by
\eqn
\hat\tr A = \bra{}A\ket{}.
\enq
It is a trace in that it satisfies the axioms for a trace, for example
$\hat\tr [A,B]=0$.  (This is not yet obvious.)

One can see graphically that the construction works,
\eqn\label{checkcon}
\hat\tr \hat M_{\alpha_1}\ldots\hat M_{\alpha_m} =
\vev{\Tr M_{\alpha_1}\ldots M_{\alpha_m}},
\enq
by expanding the product of $(l_{\alpha_i} + l_{\alpha_i}^*)$ in $2^m$ terms
and
associating a planar diagram with each non-zero term.
Consider the example in figure 4.
We start with $M_{\alpha_m=a}\ket{}$ which makes the state $\ket{a}$, and we
interpret this as adding a line with the label $a$ to the diagram.
With $M_{\alpha_{m-1}}$, there are in general two possibilities:
we can always act with $l_{\alpha_{m-1}}^*$ to create a new line labeled
$\alpha_{m-1}$, and
if $\alpha_{m-1}=\alpha_m$ we can annihilate the existing line with
$l_{\alpha_{m-1}}$ as well.
The `free' nature of the Fock space precisely reproduces the planarity
constraint on the diagrams.  After applying $M_{\alpha_1}$, we must be left
with
no lines to have a matrix element with $\bra{}$.
This proves (\ref{checkcon}), and since $\Tr$ is a trace, so is $\hat\tr$.

\includefigures\begin{figure}[bt]
\epsfxsize=9cm \epsfysize=5cm
\epsfbox{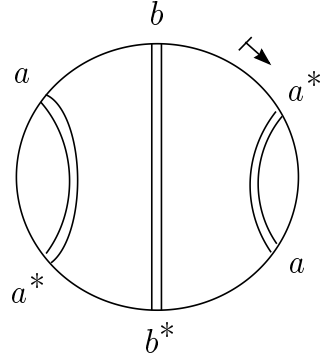}
\caption{Each non-zero term in the expansion of
$\vev{\hat\tr \hat M_b \hat M_a^2 \hat M_b\hat M_a^2 }$
corresponds to a planar diagram.}
\end{figure}\fi

The diagrammatic argument is a good demonstration but to gain insight from the
use of master fields we need new concepts.
The central concept in Voiculescu's work is that of `free probability
distribution.'
This is analogous to the idea of independence in probability theory:
a set $\{x_\alpha\}$ of commuting random variables is independent if the
joint expectations $\phi(f_1(x_1)\ldots f_n(x_n)) = \phi_1(f_1(x_1))\ldots
\phi_n(f_n(x_n))$.

\def\ex#1{\phi({#1})}
Let us regard our matrix model expectation values $\vev{\tr W}$ as giving the
joint expectations $\phi(W)$ for words constructed from a set $\{M_\alpha\}$ of
non-commuting random variables.  We normalize $\phi(1)=1$.
The distribution is {\it free} if
\eqa\label{freeas}
\phi(~f_i(M_{\alpha_i})~) = 0 \qquad i=1,2,\ldots m~~ {\rm and}\\
\alpha_i\ne\alpha_{i+1}~~\forall i \qquad\no
\ena
imply
\eqn\label{freeness}
\phi(~ f_1(M_{\alpha_1}) ~\ldots~ f_m(M_{\alpha_m}) ~) = 0.
\enq
Just as for independence, given the individual distributions this assumption
completely determines the joint distribution.
The general expectation
$\phi(~f_1(M_{\alpha_1}) ~\ldots~ f_m(M_{\alpha_m})~)$ can be computed by
induction.
Let $g_i(M_{\alpha_i}) =  f_i(M_{\alpha_i}) - \ex{ f_i(M_{\alpha_i})}$
so that $\ex{g_i(M_{\alpha_i})} = 0$;
then by substituting
$f_i = \ex{ f_i} + g_i$ for all $i$, distributing and using (\ref{freeness})
on $\phi(g_1\ldots g_m)$, we can reduce it to a sum of terms with fewer
components.

Expectations of the large $N$ limit of decoupled matrix integrals with
(\ref{decoup}) are free.  This can be proven diagrammatically, by induction on
the number of components: given (\ref{freeas}), to get a non-zero expectation
we must take diagrams with a line connecting two components, and this splits
the diagram into a product of expectations with fewer components.
It has been proven in \cite{voi} for a broader class of decoupled measures.

Freeness can be implemented directly in the master field: given the individual
master fields for the individual matrix integrals, which we already know, the
joint master field is their `free product.'  This can be defined algebraically
using the ideas we already described: the essential point is that $\hat M_i$
and $\hat M_j$ satisfy no relations.

To fit this into a larger context, we will change our language slightly:
we should think of the master field as a representation of the abstract algebra
generated by the $M_\alpha$'s, and $\ex{A}$ as a linear functional on this
algebra.  Many of the usual ideas of representation theory, such as unitary
equivalence, are directly relevant here.  As we will see later, this language
is also well suited to the gauge theory case.

The relevant algebra for a hermitian matrix integral is the algebra of
functions of a single real variable $\lambda$.  For mathematical purposes of
course we would want to be more precise.
I will not try to be so precise, but at least for our cultural benefit let us
look at the algebras which come up.
For a one-matrix distribution, $M$ is represented by the operator of
multiplication by $\lambda$ (its spectral parameter), and
$\ex{f}=\int d\lambda~\rho(\lambda) f(\lambda)$.
This only sees a subset of $\BR$, the spectrum of $M$ (support of $\rho$), so
we could equally well use the algebra of functions on ${\rm Spec} M$.
One natural object to consider is the algebra of bounded continuous functions
on ${\rm Spec} M$.
This can be made into a $C^\ast$-algebra by using the sup norm.
(For the defining axioms of a $C^\ast$-algebra, see for example \cite{murphy}.)
$C^\ast$-algebras are the most tractable general class of algebras, and this is
quite useful.  It is a theorem that every commutative $C^\ast$-algebra is an
algebra of bounded continuous functions on some space $M$, and one can recover
the topology of $M$ from the algebra $C^0(M)$.  In matrix models, it can be
relevant to the physics to know the number of components of ${\rm Spec} M$
(e.g. one-cut v.s. two-cut matrix models) and their topology, and perhaps
analogous concepts of the `non-commutative topology' of the algebras arising in
multi-matrix models or matrix field theories will eventually be relevant.

Since we have a measure, another natural algebra to consider is that of the
$L^\infty$ functions, bounded in the norm
$||f||_\infty=\lim_{p\rightarrow\infty} \left(\int d\lambda~\rho(\lambda)
|f(\lambda)|^p~\right)^{1/p}$.
This is both a $C^\ast$-algebra and a von Neumann algebra.  These functions
need not be continuous and this algebra retains very little information about
the topology of the underlying space.

Usually for us ${\rm Spec} M$ is a bounded set $I$, so we can restrict
attention to $C^0(I)$ or $L^\infty(I)$. This last is isomorphic as an abstract
algebra to $L^\infty(S^1)$, and by Fourier transform to $L(\BZ)$, the group
algebra of $\BZ$.
\footnote{The group algebra of $G$, $C(G)$, is the algebra of linear
combinations of
group elements $\sum_{g\in G} a(g) g$ with multiplication induced from the
group law, and $g^\ast=g^{-1}$.  It is essentially the convolution algebra of
functions on $G$.  Perhaps the simplest precise version of this is to take all
$L^1$ functions on $G$ (with an invariant measure).
For `nice' groups such as compact Lie groups and abelian groups, the
representation theory can be used to turn convolution into multiplication on a
dual space, as we were doing in section 2, and for the group $\BZ$, this is
just Fourier transform -- it is a good exercise to check this if you haven't
seen it.}
Now we can appeal to a result of \cite{voi} which I hope will sound intuitive:
the free product of group algebras is the algebra of the free product of
groups.  The free product $\BZ*\BZ\ldots$ ($n$ times) is the free group on $n$
generators, $F_n$.  Thus the von Neumann algebra which arises in an $n$-matrix
problem is $L(F_n)$.

It is conceivable that other algebras could come up.
For example, in the large $N$ limit of Chern-Simons theory, \cite{dcsw}
one expects Wilson loop expectations on $\Sigma\times I$ (for $\Sigma$ a
Riemann surface) to give a functional on $L(\pi_1(\Sigma))$.
In general, we know that we will get a `type ${\rm II}_1$ algebra in the
Murray-von
Neumann classification.'  This just means that there is a trace taking values
(on positive operators of norm one) in $[0,1]$.  One also has a notion of
`factor,' an algebra with trivial center, which implies that this trace is
unique.  This may sound attractive for our purposes but at this point one
starts getting into subtleties, in particular involving the choice of norm used
in defining the algebra. There are many definitions of $C(F_n)$ (for $n\ge 2$):
some are factors (e.g. the `reduced' algebra), and others are not (e.g. the
`full' or `universal' algebra).
The only point I will take from this is the obvious one that one needs to be
careful to check that one's definitions are appropriate.

Returning to a single matrix,
another description of the same representation which makes contact with our
free Fock space is to write a series
\eqn\label{matseries}
\hat M_\alpha = l_\alpha + \sum_{n\ge 0} c_n l^{*n}_\alpha
\enq
satisfying $\hat\tr \hat M_\alpha^k = \int d\lambda \rho(\lambda) \lambda^k.$
Clearly such a thing exists in general, because the $k$'th moment is determined
by $c_n$ with $n\le k-1$.
The free product of these representations can then be taken by letting these
operators act on the free Fock space.

One can show \cite{voi} that the $c_n$'s are determined as follows:
given the resolvent $F(z)=\ex{1\over z-M}=\int{d\lambda\rho(\lambda)\over
z-\lambda}$,
compute the inverse of $F(z)$ under composition, $K(F(z))=z$.  Then
\eqn\label{defK}
K(x) \equiv {1\over x} + \CR(x) = {1\over x} + \sum_{n\ge 0} c_n x^n
\enq
For example, the semicircle distribution has $F(z)={2\over
r^2}(z-\sqrt{z^2-r^2})$ giving
$\CR(x)={r^2\over 4}x$.

Thus, the operators (\ref{matseries}) acting on free Fock space are the master
field for any decoupled set of matrix integrals.
What can we do with this?
A problem treated in \cite{voi} is the following:
find the moments $\ex{(\sum_\alpha M_\alpha)^n}$, or equivalently the spectral
density for $\sum_\alpha M_\alpha$.
In ordinary probability theory, the resulting distribution would be the
convolution of the independent distributions; thus the new concept can be
called `free convolution' (or `additive free convolution').

It is clear that if we add two free variables, the result will be free with
respect to the rest, and the addition is commutative and associative.  Thus we
need only consider binary free convolution.
The result which determines it is the following:
the operators
\eqn
\hat M_1 + \hat M_2 = l_1 + l_2 + \sum c_{1,n} l_1^{*n} + c_{2,n} l_2^{*n}
\enq
have the same distribution as
\eqn
\hat M_{1+2} = l + \sum (c_{1,n}+c_{2,n}) l^{*n}.
\enq
This is not hard to see, by explicitly writing out a term like
$\vev{|(\hat M_1 + \hat M_2)^k|} = \vev{|(l_1 + l_2 + \ldots)^k|}$
and noting that any place an annihilation operator appears,
exactly one of $l_1$ or $l_2$ will contribute.
Thus
\eqn\label{freeadd}
\CR_{1+2}(x) = \CR_1(x) + \CR_2(x).
\enq
For example, the free convolution of semicircles with radius $r_i$ is
a semicircle of radius $r^2=\sum r_i^2$.

In \cite{voi} an analogy is developed between the Gaussian in ordinary
probability theory and the semicircle law in free probability theory.
For example, there is an analog of the central limit theorem:
given a free family of distributions $\phi_i$, with $\ex{M_i}=0$, all higher
moments bounded, and
$\lim_{n\rightarrow\infty} {1\over n}\sum_{i=1}^n \phi(M_i^2) =r^2/4 >0$,
the sum ${1\over\sqrt{n}}\sum_i M_i$ converges on a semicircle with radius $r$.
Such results might help explain the ubiquity of the GOE and GUE in statistics
of energy levels of random Hamiltonians.

So far we have been considering free products of one matrix distributions but
the concept is more general: for example, we could take the free product of
$\phi$ with arbitrary dependence on $M_1$ and $M_2$, with $\psi$ depending on
$M_3$.  In full generality we could talk about a free family of subalgebras
$A_i$ of the full algebra of ordered products: in (\ref{freeness}), the
successive components would be required to belong to distinct subalgebras.

Let us use these techniques
to give another derivation of (\ref{euler}),
here for quantum mechanics of a hermitian (rather than unitary) matrix.
(This is very similar to \cite{voiinf}).
We start with a discretized path integral:
\eqn
Z = \int \prod_{i=1}^L dM(i)~e^{-N\sum_i \Tr(M(i+1)-M(i))^2/\epsilon}
\enq
with $t=\epsilon L$.
Given an initial condition $M(0)$, this determines
\eqn\label{stproc}
M(t) = M(0) + \epsilon\sum_{i=1}^L \eta(i)
\enq
where the $\eta(i)=M(i)-M(i-1)$ are free random variables with semicircular
distribution.
Let $\rho_t(\lambda)$ be the resulting spectral density at time $t$, and
define $F_t$ and $K_t$ from it as above.
We can compute $K_t$ using (\ref{freeadd}) in terms of the initial condition
$K_0(u)$:
\eqn\label{stoch}
K_{t}(u) = K_{0}(u) + \epsilon L u
\enq
Changing variables from $u$ to $\xi=K_{0}(u)$,
\eqn\label{intermed}
K_t(F_0(\xi)) = \xi + t F_0(\xi)
\enq

Let us define $z(\xi,t)=K_t(F_0(\xi))$.
Applying $F_t(z)$ to both sides of this definition, we have
\eqn
F_t(z)=F_0(\xi)
\enq
and we have reproduced (\ref{hopfsol}) with $F_t(z)=p_+(z,t)$.
Finally, substituting into (\ref{intermed}) produces (\ref{hopfsoltwo}),
the general solution of (\ref{euler}).

Notice that this derivation only required an initial condition and produced the
Hopf equation directly for the resolvent
$\Tr (z-M)^{-1}$.
Perhaps we have found a better explanation for the condition
$p_\pm\sim 1/\theta$ we found in YM$_2$ on the plane?
That result was almost the same but with periodic $\theta$ -- can we derive it
in this spirit?

One can imagine calculating the master field for the holonomy in figure 5 by
summing graphs in axial gauge.
This would produce an evolution similar to (\ref{stproc}), but operating by
multiplication:
\eqn
U(t+\epsilon) = e^{iA(t,0)} U(t) e^{-iA(t,L)}
\enq
with the $\{A(t,x)\}$ at different times free with respect to each other.
This should be calculable using `multiplicative free convolution,'
another concept developed in \cite{voi}.

\includefigures\begin{figure}[bt]
\epsfbox{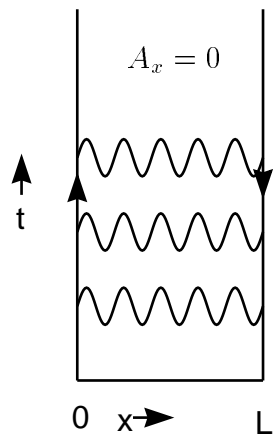}
\caption{Time evolution in YM$_2$.}
\end{figure}\fi

A direct field theory application of this is to Gaussian matrix models
(free in the physics sense).  The $D$-dimensional matrix field theory with
(Euclidean) action
\eqn\label{gaussian}
S = {N\over 2}\Tr \int d^Dx (\partial_i M)^2 + {m^2} M^2
\enq
is a free product of semicircular distributions for each mode
$A(k) + i B(k) = \int d^Dx e^{ikx} M(x)$.

It is not clear whether freeness plays any direct role in non-trivial higher
dimensional field theories, but it may be possible to use free distributions
indirectly to construct their master fields.  A construction of this type was
given by J. Greensite and M. Halpern: let us discuss it briefly.~\cite{green}

A general field theory can be formulated using stochastic quantization:
\cite{stoq}
the $\tau\rightarrow\infty$ limit of the solution of the Langevin equation
\eqn\label{lange}
{\p\over\p\tau} \hat\phi(x) = -{\p S\over \p\hat\phi(x,\tau)} + \eta(x,\tau),
\enq
where $\eta(x,\tau)$ is a Gaussian random variable with correlation
$\vev{\eta(x_1,\tau_1)\eta(x_2,\tau_2)}=2\delta(x_1-x_2)\delta(\tau_1-\tau_2)$,
reproduces quantum expectations, $\vev{\hat\phi(x)\ldots}_\eta =
\vev{\phi(x)\ldots}_\phi$.

For $\hat\phi$ a hermitian matrix, the $\eta(x,\tau)$ become free random
variables in the large $N$ limit.  (Greensite and Halpern represented them
using explicit Gaussian matrix integrals.)  Thus we can regard (\ref{lange}) as
a `free non-linear PDE' for which the $\tau\rightarrow\infty$ limit of a
solution is a master field.
This is an equation with some similarity to (\ref{stproc}), and perhaps one
could find free analogs of the existing techniques to solve non-linear PDE's
(a particularly interesting idea for integrable theories).

Let us at least solve the linear case: for (\ref{gaussian}), we can Fourier
decompose $\hat\phi=M$, and
proceeding as we did with (\ref{stproc}) for each mode, find that
\eqn
{\p\over\p\tau} K_\tau(k;u) = -(k^2+m^2)K_\tau(k;u) + u.
\enq
The solution is $K_\tau(k;u) = u/(k^2+m^2) + C u e^{-(k^2+m^2)\tau}$ which for
$\tau\rightarrow\infty$ corresponds to the semicircle we mentioned earlier.

Another possible use of free expectations would be
as ansatzes or as starting points for perturbation theory.
An example of something we can get this way is a gauge theory loop functional
which interpolates smoothly between `weak coupling' behavior of the spectral
density at high momenta and `strong coupling' behavior at low momenta.  It is
just defined by the integral
$\int dA \exp -\int d^Dk \Tr f(k,A_\mu(k)A^\mu(-k))$
(with $A_\mu(-k)=A_\mu(k)^*$).
$f(k,z)$ can be chosen arbitrarily and to get the required property
we would take it to interpolate between $f(k,z)\sim k^2 z$ at large $k$,
and (say) $f(k,z)\sim k^2 \cos \sqrt{z}$ at small $k$.

The ansatz is rather ugly, especially with its ad hoc gauge fixing.
Nevertheless it does define a master field and all Wilson loop expectations
can be computed from it.  A possible application for it would be as a `building
block' for more general loop functionals, to which we turn.

There is a more general construction, by which if one is given all expectation
values of invariant operators,
one can construct a master field which reproduces them.
This has been done by I. Singer for a continuum gauge theory, specifically
YM$_2$.~\cite{singer}
Let us discuss an analogous construction in lattice gauge theory.
(One can also adapt it for matrix models.)
In this case, the gauge holonomy is a representation of the groupoid of paths
on the lattice.%
\footnote{A groupoid is like a group, but the multiplication law need not be
defined between all pairs of elements.~\cite{connes}
Here the multiplication $p_1\circ p_2$ is defined if the end of $p_1$ is the
start of $p_2$.  In the associated algebra, a multiplication undefined in the
groupoid produces zero.}
For the present discussion (this can be generalized) we fix an `axial' gauge by
choosing a spanning tree $T$ of the lattice, and setting these links to $\bf
1$.
The holonomy is then determined by its values on the graph formed by
contracting $T$ to a point $t$; this graph is a `bouquet' of circles attached
at $t$ and paths starting and ending at $t$ generate a free group algebra.
Thus $W(L)$ is a functional on the free group, and we define its values on a
general element of the algebra by linearity.

The construction of the master field is the GNS (Gel'fand-Naimark-Segal)
construction \cite{murphy}
and once one has phrased the expectation values as a linear functional on the
group algebra, it applies straightforwardly.
The regular representation $\pi$ of the algebra $A$ acts on a vector space
$\CH\cong A$ by
\eqn
\pi(a)\ket{b} = \ket{ab}.
\enq
Essentially, we want to make this a Hilbert space with the inner product
\eqn\label{innerp}
\vev{a|b} = W(b^*a)
\enq
and define
\eqn
\Tr \pi(a) = \vev{I|\pi(a)|I}.
\enq
(We are labelling vectors by the associated element of $A$, and $I$ is the
identity element in the group.)

The master field will be the representation $\pi(L)$.
Its defining property is that $\Tr \pi(L) = W(L)$;
this is almost a tautology and in this sense we only
`repackaged' the information of the loop functional.
But there is more to check, because we need the inner product to be
positive definite, $\pi$ to be a unitary representation, and $\Tr$ to have
cyclic
symmetry.

The first condition clearly will not hold for just any $W(L)$:
it must satisfy a positivity condition.
Now positive definiteness of the inner product would follow from
$W(a^*a)>0~\forall a$, but it is clear that
semi-positive definiteness
\eqn\label{constraint}
W(a^*a) \ge 0 \qquad\forall a
\enq
is all we can expect in general.
Let's reconsider the $D=1$ case: loops are classified by winding number and
their group is $\BZ$; the positivity condition is
\eqn\label{boch}
\sum_{n,m} W_{n-m} f_n f^*_m \ge 0 \qquad \forall f_n.
\enq
We can turn the convolutions into multiplication by considering
$\rho(\theta)=\sum_n W_n e^{in\theta}$ and $f(\theta)$, in terms of which
(\ref{boch}) becomes
\eqn
\int d\theta \rho(\theta) |f(\theta)|^2 \ge 0
\enq
which is our old positivity condition.

One can define a positive element $p$ of an operator algebra as one whose
eigenvalues are positive in any representation, or more abstractly by requiring
its spectrum (this is defined without appeal to any representation as
$\{\lambda|(p-\lambda)^{-1}\notin A\}$) to be contained in the positive real
axis.
An element $p=a^*a$ will be positive, and it is a theorem for $C^\ast$ algebras
that any positive element can be written in this way.
One refers to $W$ as a positive linear functional if it is positive on the
positive elements.

We now deal with the semi-definite case by finding the ideal of elements
$\{z|W(z^*z)=0\}$ and quotienting the space $\CH$ by this ideal.
In the one matrix case, the quotient is the algebra of functions on
${\rm Supp}~ \rho$.
The inner product (\ref{innerp}) is positive definite on the quotient and there
is a unique completion of the quotient which is a Hilbert space.
The result is a unitary representation.
Unitarity $u^\ast=u^{-1}$ follows because it was true in the abstract algebra.

Finally, $\Tr [a,b] = \vev{1|[a,b]|1} = W([a,b]) = 0$.
Thus the representation $\pi(L)$ is a master field.

{\it Not} every loop functional has a master field, because of the positivity
requirement.
A (unitary) quantum field theory vacuum, however, will.
First, the contribution from any point in field space is positive: by
definition, a positive $L$ must have positive eigenvalues in any
representation, and thus $\Tr U(L)$ (for positive $L$) will be positive for any
$U$.
Then, expectation values under a functional integral with positive measure will
be positive, so these will be associated with master fields.

If we need to deal with a non-positive loop functional (say a perturbation of
the ground state), we can construct a `virtual master field' which is a pair of
fields $U_1,U_2$ such that $W(L)=\Tr U_1(L) - \Tr U_2(L)$.
It can be proven that this exists for any real $W(L)$, and that a similar
construction with four $U_i$ can reproduce any complex functional.

If we regard the large $N$ theory as being defined by a loop equation, e.g. the
Migdal-Makeenko equation or equations derived from a collective Hamiltonian, we
should regard the positivity constraint as an additional element of the
definition.
It is non-trivial, non-perturbative structure of the configuration space.
To some extent it is independent of the loop equation, and when we solve the
equation we may need to enforce the constraints by hand, as we discussed in the
previous section.\footnote{
To complete the discussion, we should prove that every positive loop functional
can be produced by some path integral and thus there are no further
constraints.  We have only done this for small modifications of the strong
coupling vacuum and this is an interesting open problem.}

These inequality constraints are rather simpler than the Mandelstam constraints
on loop functionals constructed from gauge fields at finite $N$ -- for example,
there are cases (loop functionals near the strong coupling limit) in which they
are irrelevant.
It was known in earlier work that such constraints exist and important,
\cite{jevicki,yaffetwo} and the description given there had some similarities:
when the change to loop variables is non-singular,
the metric on configuration space ($h_{ij}$ of section 3) in these variables,
\eqn\label{oldcon}
\Omega(L,L')=\sum_{L''}\Tr [E_L'',W(L)][E_L'',W(L')]
\enq
will be positive definite.
The constraint (\ref{constraint}) is a bit stronger than (\ref{oldcon}) (in
$D>2$) because it does not have the sum over $L''$.
It is simpler to think about, most importantly because it is obviously linear.
It is also clear that positivity requires positivity for any subgroup of the
group of loops.

Let us illustrate it on a `figure eight,' a bouquet of two loops $l_1$ and
$l_2$ with holonomies $U_1$ and $U_2$.  Besides a list of the $W(L)$, an
equivalent form of the data is to specify
\eqa
&W(l_1^{p_1} l_2^{q_1}\ldots l_1^{p_m} l_2^{q_m}) =\qquad\\
&\qquad\int d\theta_i d\phi_i~ \rho_m(\theta_i,\phi_i) e^{i\sum p_i\theta_i+
q_i\phi_i}.\no
\ena
The simplest new case of (\ref{constraint}) is to let
$a=\sum_{m,n} c_{m,n} l_1^m l_2^n$: these can be combined to show that
$\rho_1(\theta_1,\phi_1)\ge 0$.
We could also derive this from a spectral decomposition
$U_i = \int d\theta~\rho_i(\theta) P_i(\theta)$ over projections
$P_i(\theta)^2=P_i(\theta)$, using $\tr P~ P'\ge 0$.
However the pattern does not generalize:
the next constraints, derived from $a=\sum_{l,m,n} c_{l,m,n} l_1^l l_2^m
l_1^n$, are too weak to force $\rho_2$ to be positive.
It is not hard to construct examples in which it is not, using projections such
that $\tr P~ P'~P'' < 0$.

Although this does illustrate that (\ref{constraint}) contains new information,
this particular explicit form does not look too useful.
Another illustration of the difficulties of working explicitly with
(\ref{constraint}) is the following: there are functionals on a bouquet of $k$
loops which
satisfy the constraints defined in terms of $l<k$ loops, but violate
constraints
involving $k$ loops.

A loop functional constructed from a master field automatically satisfies the
constraints, and (as was noted in previous work) the simplest description of
the configuration space is the space of all master fields up to gauge
equivalence, or (what is the same thing) the space of unitary representations
of the free group algebra up to unitary equivalence.

An important issue in working with $D>2$ large $N$ theories is the choice
of representation of loop functionals -- for both analytical and numerical
purposes we need truncations which allow approximating a loop functional with a
finite amount of information.
If constraints are saturated, it is clear (already in $D=2$) that specifying
the functional on all loops up to a given length is inappropriate.
One alternate proposal in earlier work was to work with a finite-dimensional
approximation to the master field and derive the loop functional from it.
Another style of representation is to build loop functionals up from simple
functionals.  Since the space of loop functionals really is a linear space, it
makes sense to propose a basis for loop functionals.  One might try to find a
basis of `extreme points' in the space of functionals, from which any positive
functional could be made by superposition with positive coefficients.

Simple operations which preserve positivity are addition of loop functionals,
and loopwise multiplication $(W_1\times W_2)(L) = W_1(L) W_2(L)$.
Formally, the corresponding master fields are just the direct sum and direct
product of those for $W_1$ and $W_2$.
Although we have not precisely defined continuum master fields, it is
interesting to note that in the continuum, if $A_1$ and $A_2$ are solutions of
the classical Yang-Mills equations, both direct sum and direct product will
produce a new solution. This explains the observation of A. Migdal that the
left hand side of the Migdal-Makeenko equation (\ref{mm}) is a first order
linear operator on loop functionals.
\cite{migdal}

There is a similar idea of `phase space master field' which reproduces
expectation values of invariant operators made from $\vec A(\vec x)$ and $\vec
E(\vec x)$, or for matrix models $M(\vec x)$ and its conjugate $\Pi(\vec x)$,
as used in \cite{bardakci,yaffetwo}.
Issues of operator ordering
make this rather more subtle to study, but we intend to discuss it in future
work.

Let us discuss a simpler question raised in the last section:
what is the higher dimensional analog of the Virasoro algebra of matrix quantum
mechanics, and the corresponding group?
The generators are known:
\eqn
\Pi(f,x) = \Tr f(M)\Pi(x)
\enq
where $f(M)$ can be an arbitrary word in the $M$'s (but not the
$\Pi$'s).\footnote{
We can assume $f(M)=f(M)^+$ without loss of generality.
Strictly speaking, we should define $\Pi(f,x)=\half\Tr (f\Pi+\Pi f)$, a
self-adjoint operator, but this is not important for the present discussion.}
Their algebra is easy to compute as is their action on the invariant
observables $\phi(W)=\Tr M(x_1) \ldots M(x_m)$:
\eqa
[ \Pi(f,x), \phi(W) ] =\sum_i \delta_{x,x_i}& \\
\Tr M(x_1)\ldots M(x_{i-1})&~f~M(x_{i+1})\ldots M(x_m)\no
\ena
Can we generalize to $D>1$ the identification of this as the infinitesimal
diffeomorphism $\delta\lambda=f(\lambda)$ ?
In terms of the master field, the generalization is
\eqn
\delta M(x) = f(M)
\enq
and we should think of this transformation as an `infinitesimal free
diffeomorphism.'

It is an infinitesimal automorphism of the free group algebra $C(F_n)$:
it is linear and a derivation, and $\delta M^+=(\delta M)^+$ implies it is an
automorphism of the $C^\ast$ algebra.
Thus the corresponding group is the automorphism group of the free group
algebra (or possibly the connected component of the identity), and the
coherence group is a semidirect product of this with the additive group in the
algebra.

Returning to configuration space, a simple but important master field for gauge
theory reproduces the extreme strong coupling limit, the loop functional
$W(L)=\delta_{L,I}$.  It is just the regular representation with the
conventional inner product
$\vev{L_i|L_j}=\delta_{i,j}$ in a basis of loops.
It saturates no constraints:
by adding sources to the functional integral
we can make small variations of the $W(L)$ for
every $L$ independently.

The first observation to make is that this master field is a representation on
a very large Hilbert space, not just infinite but `more infinite' than even a
quantum field theory Hilbert space.  What I mean by this is that in terms of
the natural grading, namely by word length, the degeneracy of states grows
exponentially as $(2D-1)^l$.  (For quantum field theory in dimension $D$ and
with finitely many fields, the degeneracy of states grows as $\exp E^{1-1/D}$.)
It is the same linear space in which variations of the loop functional live and
we can (by analogy with string field theory) also think of it as the `first
quantized string Hilbert space.'
{}From this point of view, exponential
growth is the expected result for a string theory.
However there is an important difference with the better understood case of
fundamental string theory: although there we also have exponential density of
states in terms of space-time energy, the world-sheet grading is not by
space-time energy but by world-sheet `energy' $n=\alpha'm^2$, and thus
world-sheet densities of states have the familiar
$\exp \sqrt{n}$ behavior.

Now the grading which is relevant in string perturbation theory is that given
by the world-sheet Hamiltonian, and we don't know this for gauge strings.  (We
do know it in light-front quantization, at least if we call the states $\Tr O_i
\ket{0}$ `strings,' and there the world-sheet Hamiltonian determines $m^2$.)
Thus it is not yet clear if this has relevance for physics.
It turns out to be very relevant for mathematics, however.
Consider the following question: since we found that the extreme strong
coupling master field is simply the regular representation of the group
algebra, why not follow the normal procedures of group representation theory
and decompose this representation, which for compact Lie groups contains copies
of every irreducible representation?

In fact this representation is irreducible.
Many standard statements of representation theory break down for such large
groups.  Technically, free groups are not amenable, a statement whose
definition I leave to \cite{greenleaf,connes}.
Apparently, little is known about the space of all of its representations.

\section{Parting thoughts}

A lot of work on large $N$ in $D>2$ has concentrated on trying to solve models
or at least develop reliable numerical procedures for treating loop equations.
This is an attractive goal, not just for gauge theory but in my opinion even
more so for matrix models and string theory, because there the main difficulty
we face -- the large number of degrees of freedom -- is not an artifact of our
approximation but an essential feature of the problem we are studying.

We appear to be far from this goal, and I believe that a certain amount of
mathematical ground work will be required first, because the objects we are
working with are so unfamiliar.
Thus I want to conclude by proposing a simpler application of the loop space
formalisms than `solving' a large $N$ field theory:
to be the starting point for new perturbative expansions, and to better
understand the expansions we have.
These are problems which require some ability to work with the theory
non-perturbatively, may well lead to practical calculational techniques, and
could give us a better starting point for more ambitious work.

The prototype for `gauge string theory' is the strong coupling expansion.
The recent work suggests that this old formalism can be greatly improved, most
notably by formulating it in the continuum.
We need to continue it to weak coupling, but if it is qualitatively correct, as
perhaps it could be, there would be no fundamental barrier to doing this.
However, to use it one must justify not only the expansion around infinite
coupling but also the assumption that the difference between the starting loop
functional $W_0(L)=\delta_{L,I}$ and the result $W(L)$ is `small' -- not
necessarily in magnitude, but in the sense that no new structure of the
configuration space or action is required to get the correct result.
As we saw, these issues are much simpler for large $N$ gauge theories than for
finite $N$ gauge theories, but still non-trivial, and in the end this
assumption was found to be highly suspect -- there is new structure in the form
of linear constraints on physically realizable loop functionals, which is not
visible in strong coupling, and probably leads to a large $N$ transition
analogous to those in the solvable models.

We should think of `string' as a perturbative construct, describing a small
fluctuation of the fields, very analogous to particles in quantum field theory.
Thus we should not try to fix the original string construction but instead
define a new construction in which we expand around a different zeroth order
loop functional, e.g. one which saturated the right constraints.
Changing this could produce a very different string theory.
(See \cite{dcsw} for an extreme case of this.)

Another, more conventional motivation for expanding around non-trivial
configurations, is to get the physics right in the zeroth order.
A different direction would be to try to fix the obvious flaws of the weak
coupling expansion for gauge theory, such as the masslessness of the gluons, by
starting it around a non-trivial background.

To do any of this we need to better understand loop functionals and master
fields, and
I hope I have convinced the reader that there are interesting things which can
be done in these directions.  I believe this would also be invaluable for
matrix models of fundamental strings -- after all, the real point to finding a
$c>1$ matrix model is to make non-perturbative statements using it, and surely
the first step towards this is to have some non-perturbative picture of its
configuration space.  But these are problems for the future.

\medskip
It is a pleasure to acknowledge valuable discussions with M. Halpern,
V. Kazakov, I. Kostov, A. Migdal, G.
Moore, H. Neuberger, A. Polyakov, M. Staudacher, W. Taylor, and K. Woo,
and I especially thank I. Singer and D. Voiculescu for discussing their work
with me.  I also want to express my gratitude to the organizers of the Spring
School, and to the students for their interest and for their patience.

\end{document}